\documentclass[aps,prl,twocolumn,notitlepage,showpacs,superscriptaddress,am]{revtex4-1}%
\usepackage{graphicx}
\usepackage{amsmath}
\usepackage{amssymb}
\usepackage{color}
\usepackage{amsfonts}%
\providecommand{\U}[1]{\protect\rule{.1in}{.1in}}

\def\tm{(TMTTF)$_2X$}

\def\ALPHA{$\alpha$-(BEDT-TTF)$_2$I$_3$}

\def\dmit{$\beta^{\prime}$-EtMe$_3$Sb[Pd(dmit)$_2$]$_2$}
\def\EtMe{$\beta^{\prime}$-EtMe}
\def\Cu{$\kappa$-(BEDT-TTF)$_2$Cu$_2$(CN)$_3$}
\def\Ag{$\kappa$-(BEDT-TTF)$_2$Ag$_2$(CN)$_3$}

\def\Cl{$\kappa$-(BEDT-TTF)$_2$Cu[N(CN)$_2$]Cl}

\def\HgCl{$\kappa$-(BEDT-TTF)$_2$Hg(SCN)$_2$Cl}
\def\HgBr{$\kappa$-(BEDT-TTF)$_2$Hg(SCN)$_2$Br}
\def\kCu{$\kappa$-CuCN}
\def\kAg{$\kappa$-AgCN}
\def\kCl{$\kappa$-CuCl}
\def\kHgCl{$\kappa$-HgCl}
\def\kHgBr{$\kappa$-HgBr}
\def\aI3{$\alpha$-I$_3$}
\def\iToneT{$(T_1T)^{-1}$}
\def\iTone{$T_1^{-1}$}

\begin{document}
\title{Impurity Moments Conceal Low-Energy Relaxation of Quantum Spin Liquids}
\author{A. Pustogow}
\affiliation{Department of Physics and Astronomy, UCLA, Los Angeles, California 90095, USA}
\author{T. Le}
\affiliation{Department of Physics and Astronomy, UCLA, Los Angeles, California 90095, USA}
\author{H.-H. Wang}
\affiliation{Department of Physics and Astronomy, UCLA, Los Angeles, California 90095, USA}
\author{Yongkang Luo}
\affiliation{Department of Physics and Astronomy, UCLA, Los Angeles, California 90095, USA}
\author{E. Gati}
\affiliation{Institute of Physics, Goethe-University Frankfurt, 60438 Frankfurt (Main), Germany}
\author{H. Schubert}
\affiliation{Institute of Physics, Goethe-University Frankfurt, 60438 Frankfurt (Main), Germany}
\author{M. Lang}
\affiliation{Institute of Physics, Goethe-University Frankfurt, 60438 Frankfurt (Main), Germany}
\author{S. E. Brown}
\affiliation{Department of Physics and Astronomy, UCLA, Los Angeles, California 90095, USA}

\date{\today}

\begin{abstract}
We scrutinize the magnetic properties of \HgCl\ through its first-order metal-insulator transition at $T_{\rm CO}=30$~K by means of $^1$H nuclear magnetic resonance (NMR). While in the metal we find Fermi-liquid behavior with temperature-independent $(T_1T)^{-1}$, the relaxation rate exhibits a pronounced enhancement when charge order sets in. The NMR spectra remain unchanged through the transition and no magnetic order stabilizes down to 25~mK. Similar to the isostructural spin-liquid candidates \Cu\ and \Ag, $T_1^{-1}$ acquires a dominant maximum (here around 5~K). Field-dependent experiments identify the low-temperature feature as a dynamic inhomogeneity contribution that is typically dominant over the intrinsic relaxation but gets suppressed with magnetic field.
\end{abstract}

\maketitle
The rise and fall of antiferromagnetism (AFM) in correlated electron systems is intensely debated in the context of quantum spin liquids (QSL)~\cite{Balents2010,Savary2017,Zhou2017}. These elusive states of matter are expected to host exotic quasiparticles, such as neutral spinons or Majorana fermions, and have been advanced as possible platforms for quantum information applications. Following the original work of Anderson \cite{Anderson1973}, Mott insulators on frustrated lattices are considered a natural starting point for QSL realization. In this context, insulating charge-transfer salts were among the first QSL candidate systems: the compounds \Cu\ (abbreviated \kCu), \Ag\ (\kAg) and \dmit\ (\EtMe) are well described by anisotropic triangular-lattice models \cite{Kino1996,Kato2004}, and are observed to avoid long-range order to the lowest temperatures measured \cite{Shimizu2003,Itou2008}. Consequently, the nature of the ground state, as well as the factors influencing the suppression of magnetic order have been of central importance. With respect to the former, the presence of gapless fermionic excitations has been inferred from thermodynamic probes including specific heat and spin susceptibility \cite{Yamashita2008,Yamashita2011,Watanabe2012}, as well as NMR spin-lattice relaxation \cite{Shimizu2003,Itou2008,Shimizu2016}. In some cases, thermal transport and electrodynamic measurements \cite{Yamashita2010,Dressel2018,Pustogow2018spinons} have provided evidence that these gapless excitations are also mobile~\footnote{The results from Ref.~\cite{Yamashita2010} have been recently challenged in Ref.~\cite{Bourgeois-Hope2019}.}.

\begin{figure}[ptb]
\centering
\includegraphics[width=1\columnwidth]{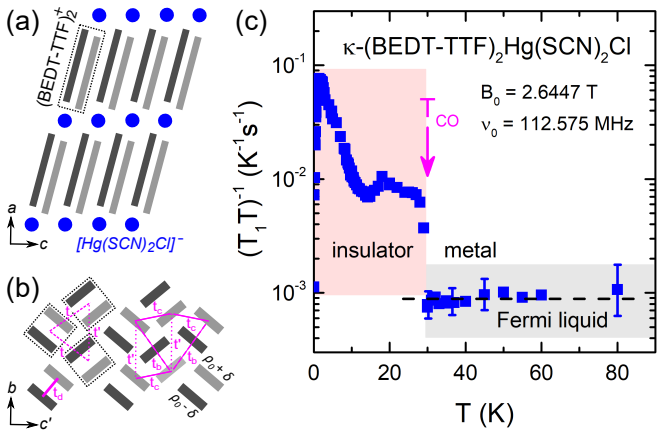}
\caption{(a) \HgCl\ crystals consist of monovalent anions (blue) separating the conducting BEDT-TTF cation layers which acquire inequivalent site charges (dark and light grey) in the charge-ordered state. (b) Dimerized in-plane arrangement with a stripe pattern of charge-rich ($\rho_0+\delta$; $\rho_0=0.5$~e) and -poor ($\rho_0-\delta$) molecules~\cite{Drichko2014,Gati2018a}. The magenta lines indicate transfer integrals $t_i$ among (BEDT-TTF)$_2^+$ dimers (black dotted lines) and between charge-rich sites, respectively~\cite{Gati2018a}. (c) In the metallic state $(T_1T)^{-1}$ is $T$-independent, in accord with Fermi-liquid behavior~\cite{Yasin2012,Drichko2014,Lohle2017}. A pronounced jump appears at the first-order MIT at $T_{\rm CO}$.
}
\label{structure}
\end{figure}

The so-called $\kappa$-phase molecular solids provide a versatile playground to study the interplay of spin and charge for varying degree of electronic correlations and geometrical frustration.
In the prototypical Mott insulators \kCu\ and \Cl\ (\kCl), pairs of BEDT-TTF molecules are strongly coupled ($t_{\rm d}\gg t$,$t'$, cf. Fig.~\ref{structure}a,b), establishing a textbook-type realization of the single-band Hubbard model at 1/2 filling~\cite{Kino1996}, even on quantitative scales~\cite{Pustogow2018}. Despite comparable exchange interaction $J/k_B\approx 200$~K, the latter compound has an AFM ground state~\cite{Miyagawa1995} while the former exhibits no magnetic order and is therefore considered as a promising QSL candidate~\cite{Shimizu2003,Shimizu2006}. Highlighting the role of frustration ~\cite{Balents2010,Kandpal2009,*Nakamura2009} in determining these disparate outcomes, despite similar structural and electronic properties, is the proposal that AFM in \kCl\ is linked to the charge degrees of freedom~\cite{Lunkenheimer2012}. That is, the detection of a dielectric anomaly~\cite{Lunkenheimer2012} and pronounced phonon renormalization effects~\cite{Matsuura2019} close to the AFM transition were assigned to intra-dimer charge degrees of freedom. It was suggested~\cite{Lunkenheimer2012} that charge order (CO) may reduce frustration giving rise to an ordered ground state. As well, quenched disorder~\cite{Guterding2015}, disorder~\cite{Huse1988,Furukawa2015b,Dressel2016,Pinteric2018,Itou2017,Lazic2018,Riedl2019}, low dimensionality~\cite{Balents2010,Powell2017}, and proximity to the Mott transition \cite{Motrunich2005} have all been cited as potentially key considerations.

A promising route to disentangle the underlying mechanisms is to introduce additional symmetry breaking. Compounds comprised of the Hg-based anions (Hg(SCN)$_2X$, $X$=Cl, Br) have recently come into focus~\cite{Yasin2012,Drichko2014,Lohle2017,Ivek2017,Gati2018a,Hemmida2018,Gati2018,Hassan2018,Hassan2019} due to the tendency towards electronic CO. The weaker dimerization (the ratios $t_d/t$ are closer to unity~\footnote{While for \HgCl\ $t_d/t'\approx 3$, the paradigmatic $\kappa$-phase materials \kCu, \kAg\ and \kCl\ have larger ratios of approximately 4--5.}) increase the relative importance of inter-site Coulomb repulsion. In \HgCl\ (\kHgCl) the metal-insulator transition (MIT) at $T_{\rm CO}=30$~K is very similar to CO in \ALPHA, also exhibiting a discontinuous symmetry breaking~\cite{Drichko2014,Lohle2017,Gati2018a,Yue2010,Ivek2011}. While the charge sector of \kHgCl\ \cite{Yasin2012,Drichko2014,Lohle2017,Hassan2019} has been investigated in great detail, no definitive conclusion was achieved on the spin degrees of freedom~\cite{Yasin2012,Gati2018a}. Particularly in view of the closely related \HgBr\ (\kHgBr), where recently an exotic dipole-liquid state~\cite{Hassan2018} and indications for ferromagnetism~\cite{Hemmida2018} were reported, the magnetic ground state and possible spin-charge coupling call for clarification.

In this Letter we investigate the low-energy magnetic properties of \HgCl\ via $^1$H nuclear magnetic resonance (NMR). In the metallic phase we observe Fermi-liquid behavior with constant $(T_1T)^{-1}$ while for 25~mK $\leq T < T_{\rm CO}$ spectroscopic measurements find no evidence for magnetic order. $T_1^{-1}$ exhibits a dominant maximum around 5~K with pronounced magnetic field and temperature dependences characteristic of $S$=1/2, $g$=2 impurity states. Notably, the overall behavior is decidedly similar to that reported for the well-known $\kappa$-phase QSL candidates, \kCu\ and \kAg. As we will argue below, it appears that the dynamic low-temperature contribution is a common feature in all these compounds without magnetic order and originates from inhomogeneities rather than intrinsic spin degrees of freedom. We quantitatively link $T_1^{-1}$ to impurity states detected by ESR~\cite{Yasin2012,Gati2018a}.

\HgCl\ single crystals with typical dimensions of $1\times 0.5\times 0.3$~mm were grown by electrochemical methods reported elsewhere \cite{Gati2018a}. NMR experiments were performed with home-built spectrometers utilizing superconducting magnets.
For sample 1, the field strength was $B_0=2.6447$~T, with alignment close to $\mathbf{B_0}\parallel c$. Field-dependent measurements (sample 2; $\mathbf{B_0}$ out-of-plane) covered the range 1.2--9.3~T.
Standard $^4$He flow cryostats were employed above 1.6~K whereas a $^3$He/$^4$He dilution refrigerator allowed us to access the range down to 25~mK. The spin-lattice relaxation rate was determined via free-induction decay following saturation, and analyzed using stretched-exponential fits.

\begin{figure}[ptb]
\centering
\includegraphics[width=0.8\columnwidth]{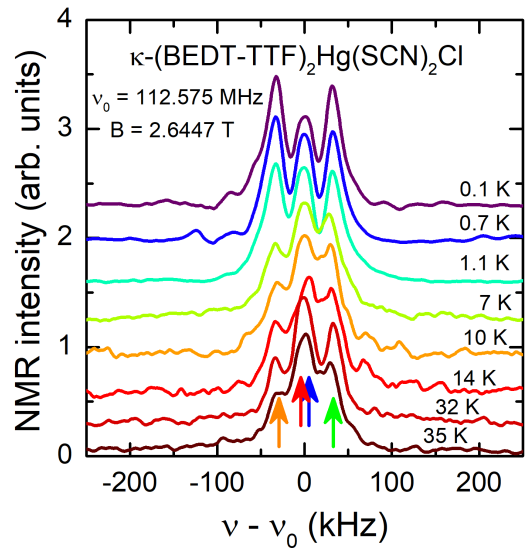}
\caption{The shape and width of the $^1$H NMR spectra remains unaffected upon cooling through $T_{\rm CO}=30$~K, ruling out magnetic order down to mK temperatures. The NMR intensity was normalized with respect to the $1/T$ enhancement; curves were shifted vertically. The minor difference in relative amplitudes of inner and outer peaks below and above 2~K is due to slightly different sample alignment in different cryostats.}
\label{spectra}
\end{figure}

The crystal structure of \kHgCl\ consists of layers of positively charged BEDT-TTF molecules separated by monovalent anions, see Fig.~\ref{structure}(a,b). Within the conducting planes the organic cations are arranged in weakly bound pairs ($t_d/t'\approx 3$) assembled in an anisotropic triangular lattice ($t'/t=0.79$~\cite{Gati2018a}), suggesting significant geometrical frustration. For $T<T_{\rm CO}$, the electronic charge is redistributed between the two sites within a dimer, likely forming a stripe-like pattern~\cite{Drichko2014,Gati2018a} that alters the magnetic frustration.
Fig.~\ref{structure}(c) shows the variation of \iToneT\ with temperature, which is $T$--independent in the metallic state ($T>T_{\rm CO}$). An abrupt jump appears at the transition signalling a change of the relevant energy scale from $E_F$ in the metal ($10^3$--$10^4$~K) to $J$ in the insulating state ($10^2$~K). The non-monotonic behavior upon further cooling will be discussed in the next paragraph.
In Fig.~\ref{spectra} we show the $^1$H NMR spectra for different temperatures, which appear to consist of four distinct peaks resulting from proton-proton dipolar coupling~\footnote{The spectrum is actually comprised of a superposition of 8 inequivalent but unresolved protons sites. See Supplemental Material for angle-dependent measurements.}. No significant modification of the peak structure is observed upon cooling below $T_{\rm CO}$ -- clearly different to AFM in \kCl\ \cite{Miyagawa1995}. Thus, the NMR spectra of \kHgCl\ show no indications of magnetic order throughout the CO phase.

\begin{figure}[tb]
\centering
\includegraphics[width=1\columnwidth]{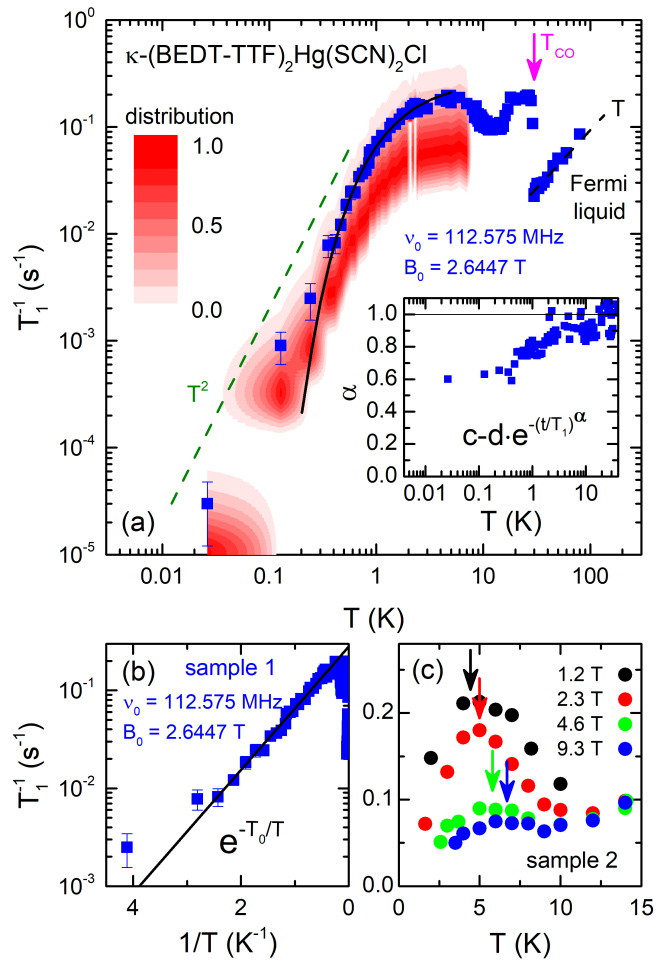}
\caption{(a)~Subsequent to the abrupt increase at $T_{\rm CO}$, the spin-lattice relaxation rate drops upon cooling, and a broad maximum forms around 5~K.
Well below 1~K $T_1^{-1}$ shows a power-law behavior similar to various spin-liquid candidates~\cite{Shimizu2003,Shimizu2006,Itou2010,Shimizu2016}. Inset: The stretched-exponential recovery ($\alpha = 0.6$ at lowest $T$) reveals a continuum of low-energy decay channels; we visualize the related distribution of $T_1^{-1}$ (according to Ref.~\onlinecite{Johnston2006}) by the red-white false-color plot in the main graph. (b) Below the peak $T_1^{-1}$ exhibits Arrhenius-like activation (black solid line; also indicated in (a)), with $k_B T_0\approx \mu_B B_0$.
(c)~Upon increasing $B_0$ the maximum is strongly suppressed and shifts to higher $T$, in excellent agreement with Eq.~\ref{dipolar-impurity-spins} -- even in the absolute values of $T_1^{-1}$.
}
\label{inverse-T1}
\end{figure}

The spin-lattice relaxation rate $T_1^{-1}$ is displayed on double-logarithmic scales in Fig.~\ref{inverse-T1}(a), covering the temperature range 0.025--80~K. For $T>T_{\rm CO}$, the relaxation process proceeds homogeneously, as evident from the single-exponential recovery ($\alpha=1$ in the stretched-exponential fit). Upon lowering $T$ within the insulating state, $T_1^{-1}$ first decreases, but then increases and peaks at $T\simeq5$~K. In this range also stretched-exponential behavior sets in (initially $\alpha\approx 0.9$, see Fig.~\ref{inverse-T1}(a) inset). Well below the maximum $T_1^{-1}$ exhibits a smooth, power-law like ($\propto T^2$) decrease on cooling further to $T\sim 25$~mK, in accord with the absence of AFM concluded from the NMR spectra (Fig.~\ref{spectra}). Stretched-exponential behavior becomes more pronounced at the lowest measured temperatures -- generally an indicator for a range of characteristic relaxation time scales. In particular, $\alpha \approx 0.6$ results from a $T_1^{-1}$ distribution spanning approximately one order of magnitude~\cite{Johnston2006}, which we illustrate by the red-white false-color plot behind the data in Fig.~\ref{inverse-T1}(a).

The low-temperature relaxation of \kHgCl\ is reminiscent of the widely studied QSL candidates \kCu, \kAg\ and \EtMe. In those cases, power-law variation with temperature has been attributed to a gapless continuum of spin excitations~\cite{Shimizu2003,Shimizu2006,Itou2010,Shimizu2016}. Here, we consider an alternative scenario: the proton \iTone\ at low temperatures is caused by dipolar coupling to localized $S=1/2$, $g=2$ spin degrees of freedom. The general idea is that the impurity spins, embedded in an otherwise nonmagnetic background, are sufficiently polarized in nonzero magnetic fields at low enough temperature, so as to progressively freeze out this relaxation channel. We note that low-temperature effects from disorder-induced spin defects were recently considered in Ref.~\cite{Riedl2019}.

\begin{figure*}[ptb]
\centering
\includegraphics[width=2.05\columnwidth]{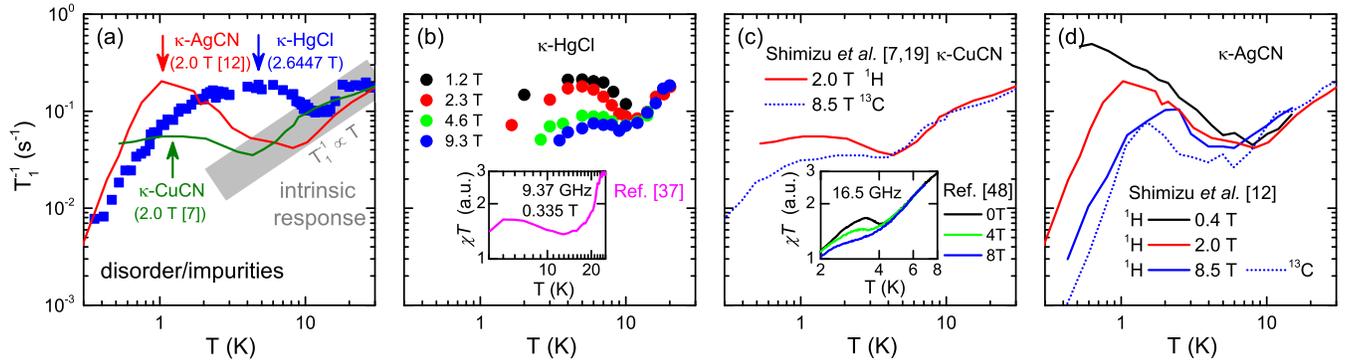}
\caption{(a) At temperatures above the maximum, the $^1$H $T_1^{-1}$ data in the insulating state of \kHgCl\ coincide with the paradigmatic QSL candidates \kCu\ and \kAg. Here, $T_1^{-1}$ follows a field-independent approximately linear $T$ dependence suggesting that this is the intrinsic response with $J\approx 200$~K. (b-d) While peaked at different $T_{max}$, the low-temperature contribution exhibits a similar suppression with higher $B_0$ for all three compounds; $^{13}$C data (scaled by $\gamma_n$ [49]) match well with the $^1$H results acquired at the same $B_0$~\cite{Shimizu2003,Shimizu2006,Shimizu2016}. A similar field-dependent contribution is observed in high-frequency susceptibility data plotted as $\chi T$ (inset of (b) at 9.37~GHz~~\cite{Gati2018a}; inset of (c) at 16.5 GHz~\cite{Poirier2012}).
}
\label{field-dependence}
\end{figure*}

The nuclear relaxation by dipolar coupling to magnetic impurities implies certain behaviors that can be compared to experiment. 
For example, $T_1^{-1}$ of \kAg\  is strongly reduced with increasing $B_0$~\cite{Shimizu2016}; similar behavior is seen for \kHgCl\ in Fig.~\ref{inverse-T1}(c). Here the field dependence is pronounced near the maxima around 5~K, while the relaxation for $T\simeq 10$~K remains rather unaffected. At a semi-quantitative level, this is precisely the temperature range corresponding to the Zeeman energy of a free spin. 
More specifically, the peak and low-temperature suppression of \iTone\ is modelled for a \textit{single} proton as
\begin{align}
T_1^{-1} = \frac{2}{5}\mu_o^2\gamma_s^2\gamma_I^2\hbar^2(S(S+1))r^{-6}\frac{\tau}{1+\omega^2\tau^2},
\label{dipolar-impurity-spins}
\end{align}
where $ 1/\tau$ is the bandwidth of longitudinal field fluctuations; it is taken to be of the form $\tau=\tau_0 e^{E_Z/k_BT}$, with $E_Z=g\mu_BSB_0$ the Zeeman energy splitting of the impurity spin levels, using $g=2$ and $S=1/2$. The activated behavior arises from the polarization of the impurity spins in the applied magnetic field. 
The dipolar coupling depends on the distance $r$ between the impurity spin and the nuclear site. Naturally, random arrangement of the former is related to a distribution of local fields which results in a stretched-exponential recovery.

Looking at the Arrhenius plot in Fig.~\ref{inverse-T1}(b), the behavior on the low-temperature side of the maximum closely follows the associated thermal activation with $k_B T_0\approx \mu_B B_0$ down to 0.2~K. The peak value in Fig.~\ref{inverse-T1}(c) roughly follows the expected $(T_1^{-1})_{max}\propto 1/B_0$ dependence, and $\tau=\omega^{-1}$ at the maximum yields $\tau_0$ in the ns range, in agreement with the ESR linewidth $\Delta H \approx 3$~mT in the insulating state~\cite{Gati2018a}.  Plugging this into Eq.~\ref{dipolar-impurity-spins}, together with our experimental values of $T_1^{-1}$, yields $r\approx 6-7$~nm. A similar result is obtained from the Curie behavior of the $T$-dependent ESR intensity~\cite{Gati2018a,Yasin2012}, giving an impurity concentration of order $10^{-2}$ per unit cell~\cite{Drichko2014}.

In Fig.~\ref{field-dependence}(a) we compare $T_1^{-1}$ of \kHgCl\ with the isostructural QSL candidates \kCu\ \cite{Shimizu2003} and \kAg\ \cite{Shimizu2016} on common scales and for comparable $B_0$ as indicated. Although at different temperatures and not necessarily of the same origin, in all these compounds we identify a dynamic contribution with similar characteristics as elaborated above for \kHgCl. Above the low-temperature maximum, 10~K $\leq T\leq 30$~K, the data are similar in magnitude and follow an approximately linear temperature dependence; in the case of \kCu\ and \kAg, the behavior is attributed to gapless spinons. Generally, however, the quantitative similarity across compounds is not surprising in view of the comparable exchange energies. Since the dynamic maximum dominates a large range of the low-temperature relaxation, we cannot conclude whether there is a spin gap or not. High-field experiments ($k_B T_{max}<\mu_B B_0<J$, i.e. a few tens of T) could possibly disclose the intrinsic magnetic properties of the QSL candidates down to low temperatures.

The overall suppression of the $g=2$, $S=1/2$ peak with increasing $B_0$ is similar for \kHgCl\ and \kAg, as summarized in Fig.~\ref{field-dependence}(b,d). The published  $T_1^{-1}$~\cite{Shimizu2016} on $^1$H and $^{13}$C~\footnote{Upon proper renormalization to the gyromagnetic ratios and the local charge density, cf. Fig.~4(a) in Ref.~\onlinecite{Shimizu2016}, the $^1$H and $^{13}$C data coincide in the relevant temperature range ($k_B T\leq J$) when acquired at the same magnetic field. Upon changing $B_0$, the apparently intrinsic response at temperatures above the maximum remains unaltered, while the low-$T$ behavior exhibits pronounced modifications, including suppression by field \cite{Shimizu2016}.} consistently show pronounced field dependence around the maximum, while the intrinsic response at higher $T$ remains unaffected. A similar feature is also seen in the magnetic susceptibility: in the insets of (b,c) we show $\chi T$ in order to compare to $T_1^{-1}$~\cite{Gati2018a,Poirier2012}. Similar to \kHgCl\ and \kAg, the $^1$H and $^{13}$C data of \kCu\ aquired at 2 and 8.5~T~\cite{Shimizu2003,Shimizu2006}, respectively, coincide above 4~K but deviate around the bump at lower $T$ [Fig.~\ref{field-dependence}(c)], where appreciable field dependence is also seen by different probes~\cite{Poirier2012,Pratt2011,Isono2016,*Isono2018}. Due to the lack of consistent $T_1^{-1}(T)$ data upon varying $B_0$, we do not exclude other contributions below 4~K in \kCu.

Even though the NMR characteristics of \kHgCl\ resemble the response of various QSL candidates in minute detail, its thermodynamic properties clearly indicate the absence of itinerant spin and charge excitations. That is, extrapolating $C/T$ down to $T=0$ yields a Sommerfeld coefficient non-distinguishable from zero~\cite{Hassan2018}, at least much smaller than for \kCu\ and \kAg\ where $\gamma\approx 10$--20~mJK$^{-2}$mol$^{-1}$~\cite{Yamashita2008,Shimizu2016}. Note, the sister compound \kHgBr, where fluctuating CO has been suggested~\cite{Hassan2018}, exhibits $\gamma$ comparable to the QSL candidates. Thus, the reduced entropy in \kHgCl\ is consistent with gapped charge and spin degrees of freedom, for instance like in a valence bond solid.
Similar to \kCu\ \cite{Yamashita2008}, $C/T$ from Ref.~\onlinecite{Hassan2018} reveals a Schottky-like increase towards lower temperatures setting in at a few 100~mK, coincident with the power law in $T_1^{-1}$. It remains to elucidate to what extent disorder is relevant for the material under study -- in particular in view of the stretched-exponential relaxation at low temperatures that suggests a continuum of low-energy decay channels.

The absolute values and temperature of the maximum in $T_1^{-1}$ differ from compound to compound. If the origins were similar, this could be associated with a varying distribution of time scales $\tau$. Performing a similar dipolar relaxation analysis for \kCu\ and \kAg\ yields slightly lower impurity densities than in \kHgCl, but of similar order of magnitude (see Supplement).
Finally, we comment briefly on the origin of the magnetic impurities in \kHgCl. The clearly discontinuous phase transition at 30 K allows for the possibility of multiple CO domains and accompanying domain walls, as recently observed in \tm\ by Raman spectroscopy \cite{Swietlik2017}. A possible scenario is that the impurity states are located at domain walls. If that were the case, the absence of CO in \kCu\ and \kAg\ would point to a different origin of the dynamic contribution, likely linked to the anion layers~\cite{Padmalekha2015,Dressel2016,Pinteric2018}. Further, recent Raman experiments on \kHgCl\ suggest BEDT-TTF$^{+0.5}$ below 10~K~\cite{Hassan2019} which could also provide a source of $g=2$, $S=1/2$ spins.

To summarize, we map the low-energy spin dynamics in \HgCl\ through the metal-insulator transition by comprehensive $^1$H NMR experiments. The spin-lattice relaxation rate indicates a Fermi-liquid metal at elevated temperatures, and exhibits a pronounced discontinuous increase upon cooling through $T_{\rm CO}=30$~K into the charge-ordered phase. From the unaltered NMR spectra (Fig.~\ref{spectra}) and the smooth temperature dependence of $T_1^{-1}$ upon $T\rightarrow 0$ (Fig.~\ref{inverse-T1}), we conclude the absence of long-range magnetic order. Notably, we find that the magnetic response is essentially identical to isostructural QSL candidates~\cite{Shimizu2003,Shimizu2006,Shimizu2016}, including the stretched-exponential recovery and a power-law like tail well below 1~K as well as a pronounced maximum in $T_1^{-1}$ (peaked around 5~K in \kHgCl). This low-$T$ contribution exhibits a strong field dependence, very similar for \kHgCl\ and \kAg, likely originating from dipolar coupling to impurity spins. Taken together, these results imply that the low-temperature NMR properties in all these frustrated materials~\cite{Shimizu2003,Shimizu2006,Itou2010,Shimizu2016} are dominated by extrinsic magnetic contributions. Suppressing the dynamic relaxation channels with high fields ($B_0\geq 10$~T), in principle, recovers the intrinsic electronic response, providing a promising route to answer the question about a spin gap in the triangular systems. Given the lack of a non-zero fermionic contribution to the low-temperature specific heat \cite{Hassan2018}, the case for a gapped ground state is stronger for \kHgCl\ than it is for \kCu\ and \kAg.

\acknowledgments We thank N. Drichko, K. Kanoda, R. Valent\'i, S. Winter, M. Dressel and A.-M. Tremblay for useful comments and discussions. A. P. acknowledges support by the Alexander von Humboldt Foundation through the Feodor Lynen Fellowship. This work was supported by the National Science Foundation (DMR-1709304). Work performed in Frankfurt was supported by the Deutsche Forschungsgemeinschaft through the Transregional Collaborative Research Center SFB/TR49.

\newpage
\cleardoublepage

\onecolumngrid
\appendix

\linespread{1.25}
\section*{Supplemental Material}

\setcounter{table}{0}
\setcounter{figure}{0}
\setcounter{equation}{0}
\renewcommand{\thefigure}{S\arabic{figure}}
\renewcommand{\thetable}{S\arabic{table}}
\renewcommand{\theequation}{S\arabic{equation}}

\section{Impurity Density Estimate}

\begin{figure}[b]
\centering
\includegraphics[width=0.6\columnwidth]{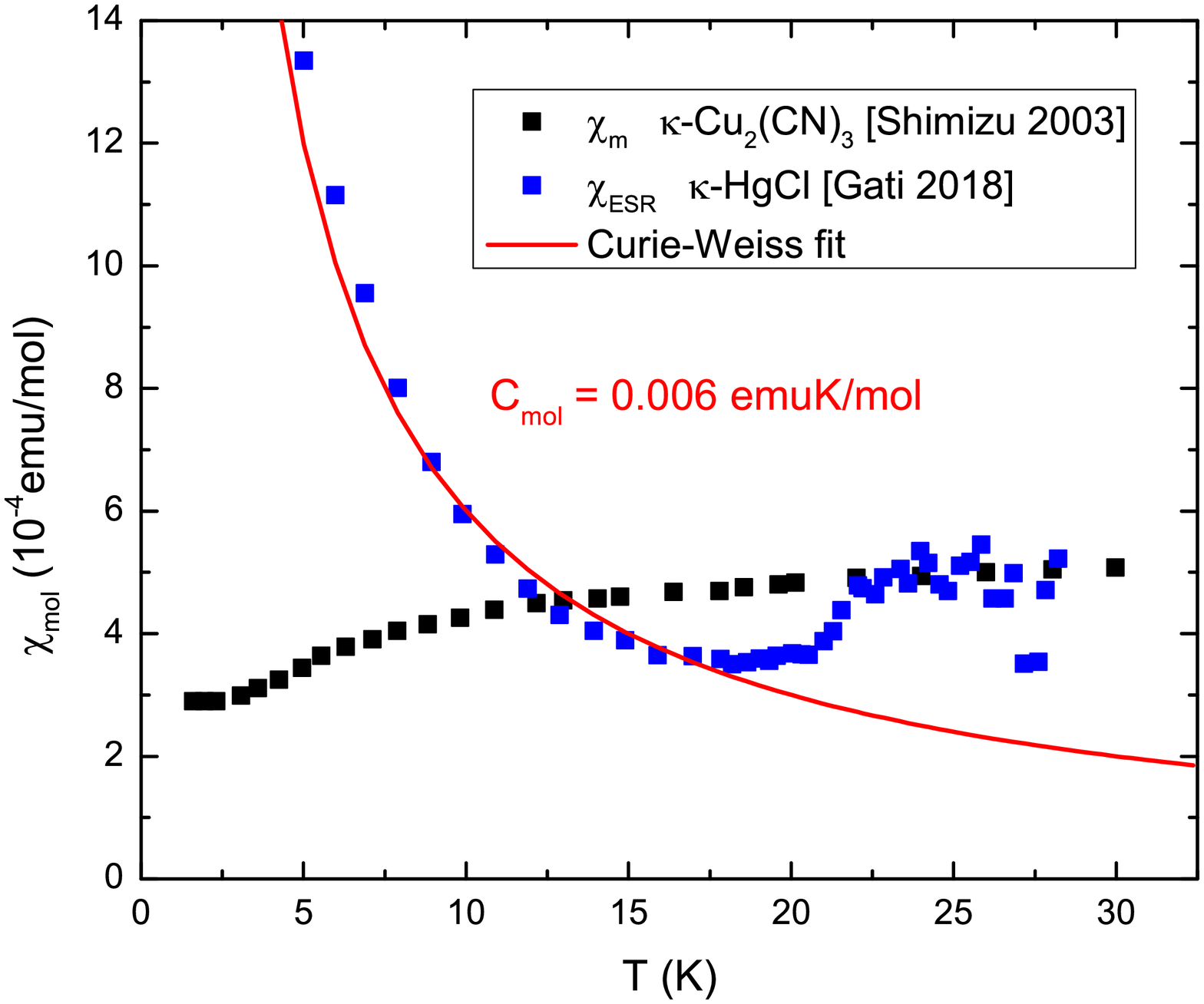}
\caption{The absolute values of the spin susceptibility of \HgCl\ (determined by ESR measurements in Ref.~\cite{SGati2018a}) was approximated by $\chi_{mol}$ of \Cu\ \cite{SShimizu2003}.
}
\label{Curie-const}
\end{figure}

The impurity spin density, $N$, of \HgCl\ was estimated by analyzing the Curie-like region of the spin susceptibility below 15~K. In order to assign absolute values to the ESR data from Ref.~\cite{SGati2018a} (Fig. S7 in Supplementary Materials), we scaled the ESR intensity to match typical values of $\kappa$-compounds in the range above the maximum, i.e. $10 \;{\rm K}\leq T\leq 30$~K. In Fig.~\ref{Curie-const} we plot $\chi_{ESR}$ and scale it to match with the susceptibility of \Cu\ \cite{SShimizu2003}, which has similar structural properties and exchange energy. The Curie-Weiss fit below 15~K yields a molar Curie constant $C_{mol}\approx 0.006$ emuK/mol. From the Curie law
\begin{equation}
    C = \frac{\mu_o\mu_B^2}{3k_B}Ng^2(S(S+1))=\chi T
\end{equation}
we obtain the relation
\begin{equation}
    N = \frac{k_B C_{mol}}{\mu_o\mu_B^2 N_A}
\end{equation}
where $N$ is the number of impurities per unit cell and we used the values $S=1/2$, $g = 2$ and $V_{UC} = 3500 $\AA$^3$ \cite{SDrichko2014}. Plugging in the above determined value of $C_{mol}$, we obtain $N = 0.016$ spins/unit cell. This impurity spin density corresponds to an average distance between protons and impurities on the order of a few nm.

Since for paramagnets $\chi T  \propto T_{1}^{-1}$, it is reasonable to use Fig. 4(c,d) to extrapolate the impurity spin densities for \Cu\ and \Ag\ from their $T_{1}^{-1}$, using the same $C$. The results of this calculation are shown in Table \ref{tab:1}.
\begin{table}[t]
\begin{tabular}{|l|c|c|c|c|}
\hline
\multicolumn{1}{|c|}{Compound} & \multicolumn{1}{l|}{$1/T_{1}$ (1/s)} & \multicolumn{1}{l|}{$C$ (emuK/mol)} & $V_{UC}$ (\AA$^3$) & \multicolumn{1}{l|}{$N$ } \\ \hline
$\kappa$-Hg-Cl                 & 0.1                                            & 0.006                              & 3500 \cite{SDrichko2014}                     & 0.016                                  \\
$\kappa$-Ag-CN                 & 0.04                                           & 0.0024                             & 1756 \cite{SShimizu2016}                     & 0.0064                                 \\
$\kappa$-Cu-CN                 & 0.03                                           & 0.0018                             & 1695 \cite{SJeschke2012}                     & 0.0048                                \\ \hline
\end{tabular}
\caption{Relaxation rate at the maximum, Curie constant, unit cell volume and resulting impurity spin density $N$ (per unit cell) for \HgCl, \Cu\ \cite{SJeschke2012}, and \Ag\ \cite{SShimizu2016,SHiramatsu2017}.}
\label{tab:1}
\end{table}

As an independent check, we now estimate the average distance between $^1$H nuclear spins in \kHgCl\ and the impurity spins based on Eq.~(1). For that, we approximated $\tau$ from the ESR linewidth in the insulating state \cite{SGati2018a}.  For $\Delta H = 30$ G, uncertainty principle yields a mutual spin flip rate of $\Delta t = \tau = 9.5*10^{-10}$ s.
Solving Eq.~(1) \cite{SAbragam1983} in the low-frequency limit, using $T_1 = 10$ s, we obtained $r = 7$ nm, which is comparable to the approximation above obtained from the impurity density $N$ via the Curie constant $C$. \textit{Vice versa}, this means that there is even quantitative agreement (within a factor two, given the approximations made) of the experimentally observed $T_1$ with the discussed model of dipolar coupling between protons and impurity spins.

\section{NMR Spectra Upon Rotation}

\begin{figure}[h]
\centering
\includegraphics[width=0.6\columnwidth]{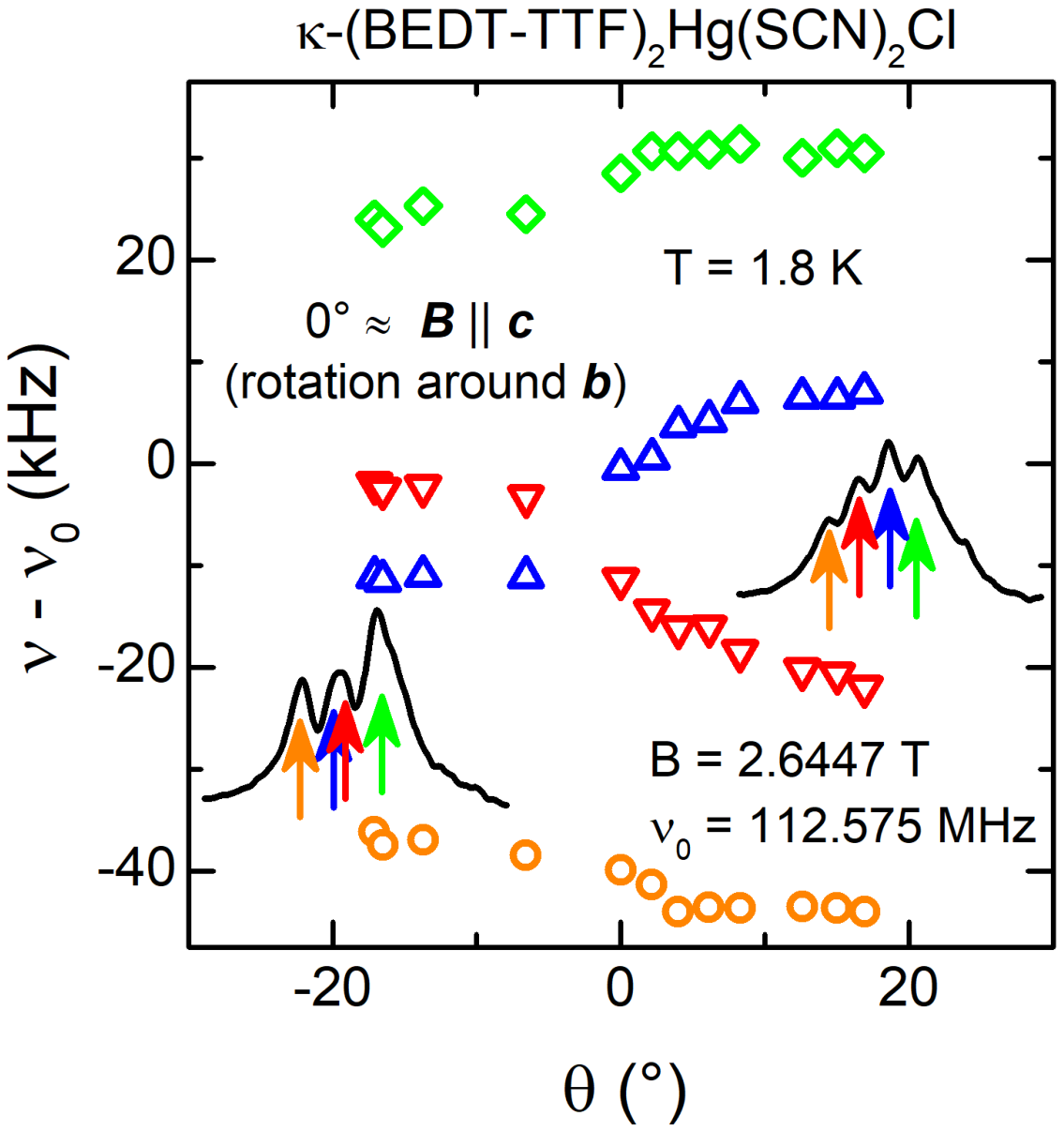}
\caption{NMR spectra upon rotation reveal four main modes. The temperature-dependent measurements presented in the main manuscript were performed at $\theta=0^{\circ}$, which is close to the crystallographic $c$-axis.
}
\label{rotation}
\end{figure}

In order to evaluate the $^1$H NMR spectra, we performed angle-dependent measurements at 1.8~K using a piezoelectric rotator. While the line shape in Fig.~2 of the main manuscript shows a trident-like structure, Fig.~\ref{rotation} reveals four main peaks that follow a common angle dependence. At the angle where temperature-dependent experiments were carried out ($0^{\circ}$) the two inner peaks are simply too close together to be distinguished. Given this proximity of the peaks, and generally reduced signal-noise ratio at elevated temperatures, we did not observe any changes in the spectrum upon crossing the glass transition at $T_g = 63$~K (cooling rate 0.5~K/min). As the focus of the present study was on low temperatures, the limited number of $T_1$ data points around $T_g$ does not capture the glass transition. Future work could target on this interesting variant, where only one (of the two) ethylene endgroups is involved~\cite{SGati2018}, which constitutes a potential source of disorder in the system, e.g. dependent on the cooling rate through $T_g$.

\begin{thebibliography}{56}%
\makeatletter
\providecommand \@ifxundefined [1]{%
 \@ifx{#1\undefined}
}%
\providecommand \@ifnum [1]{%
 \ifnum #1\expandafter \@firstoftwo
 \else \expandafter \@secondoftwo
 \fi
}%
\providecommand \@ifx [1]{%
 \ifx #1\expandafter \@firstoftwo
 \else \expandafter \@secondoftwo
 \fi
}%
\providecommand \natexlab [1]{#1}%
\providecommand \enquote  [1]{``#1''}%
\providecommand \bibnamefont  [1]{#1}%
\providecommand \bibfnamefont [1]{#1}%
\providecommand \citenamefont [1]{#1}%
\providecommand \href@noop [0]{\@secondoftwo}%
\providecommand \href [0]{\begingroup \@sanitize@url \@href}%
\providecommand \@href[1]{\@@startlink{#1}\@@href}%
\providecommand \@@href[1]{\endgroup#1\@@endlink}%
\providecommand \@sanitize@url [0]{\catcode `\\12\catcode `\$12\catcode
  `\&12\catcode `\#12\catcode `\^12\catcode `\_12\catcode `\%12\relax}%
\providecommand \@@startlink[1]{}%
\providecommand \@@endlink[0]{}%
\providecommand \url  [0]{\begingroup\@sanitize@url \@url }%
\providecommand \@url [1]{\endgroup\@href {#1}{\urlprefix }}%
\providecommand \urlprefix  [0]{URL }%
\providecommand \Eprint [0]{\href }%
\providecommand \doibase [0]{http://dx.doi.org/}%
\providecommand \selectlanguage [0]{\@gobble}%
\providecommand \bibinfo  [0]{\@secondoftwo}%
\providecommand \bibfield  [0]{\@secondoftwo}%
\providecommand \translation [1]{[#1]}%
\providecommand \BibitemOpen [0]{}%
\providecommand \bibitemStop [0]{}%
\providecommand \bibitemNoStop [0]{.\EOS\space}%
\providecommand \EOS [0]{\spacefactor3000\relax}%
\providecommand \BibitemShut  [1]{\csname bibitem#1\endcsname}%
\let\auto@bib@innerbib\@empty
\bibitem [{\citenamefont {Balents}(2010)}]{Balents2010}%
  \BibitemOpen
  \bibfield  {author} {\bibinfo {author} {\bibfnamefont {L.}~\bibnamefont
  {Balents}},\ }\href {\doibase 10.1038/nature08917} {\bibfield  {journal}
  {\bibinfo  {journal} {Nature}\ }\textbf {\bibinfo {volume} {464}},\ \bibinfo
  {pages} {199} (\bibinfo {year} {2010})}\BibitemShut {NoStop}%
\bibitem [{\citenamefont {Savary}\ and\ \citenamefont
  {Balents}(2017)}]{Savary2017}%
  \BibitemOpen
  \bibfield  {author} {\bibinfo {author} {\bibfnamefont {L.}~\bibnamefont
  {Savary}}\ and\ \bibinfo {author} {\bibfnamefont {L.}~\bibnamefont
  {Balents}},\ }\href {\doibase 10.1088/0034-4885/80/1/016502} {\bibfield
  {journal} {\bibinfo  {journal} {Rep. Prog. Phys.}\ }\textbf {\bibinfo
  {volume} {80}},\ \bibinfo {pages} {016502} (\bibinfo {year}
  {2017})}\BibitemShut {NoStop}%
\bibitem [{\citenamefont {Zhou}\ \emph {et~al.}(2017)\citenamefont {Zhou},
  \citenamefont {Kanoda},\ and\ \citenamefont {Ng}}]{Zhou2017}%
  \BibitemOpen
  \bibfield  {author} {\bibinfo {author} {\bibfnamefont {Y.}~\bibnamefont
  {Zhou}}, \bibinfo {author} {\bibfnamefont {K.}~\bibnamefont {Kanoda}}, \ and\
  \bibinfo {author} {\bibfnamefont {T.-K.}\ \bibnamefont {Ng}},\ }\href
  {https://link.aps.org/doi/10.1103/RevModPhys.89.025003} {\bibfield  {journal}
  {\bibinfo  {journal} {Rev. Mod. Phys.}\ }\textbf {\bibinfo {volume} {89}},\
  \bibinfo {pages} {25003} (\bibinfo {year} {2017})}\BibitemShut {NoStop}%
\bibitem [{\citenamefont {Anderson}(1973)}]{Anderson1973}%
  \BibitemOpen
  \bibfield  {author} {\bibinfo {author} {\bibfnamefont {P.~W.}\ \bibnamefont
  {Anderson}},\ }\href {\doibase
  http://dx.doi.org/10.1016/0025-5408(73)90167-0} {\bibfield  {journal}
  {\bibinfo  {journal} {Mater. Res. Bull.}\ }\textbf {\bibinfo {volume} {8}},\
  \bibinfo {pages} {153} (\bibinfo {year} {1973})}\BibitemShut {NoStop}%
\bibitem [{\citenamefont {Kino}\ and\ \citenamefont
  {Fukuyama}(1996)}]{Kino1996}%
  \BibitemOpen
  \bibfield  {author} {\bibinfo {author} {\bibfnamefont {H.}~\bibnamefont
  {Kino}}\ and\ \bibinfo {author} {\bibfnamefont {H.}~\bibnamefont
  {Fukuyama}},\ }\href {\doibase 10.1143/JPSJ.65.2158} {\bibfield  {journal}
  {\bibinfo  {journal} {J. Phys. Soc. Jpn.}\ }\textbf {\bibinfo {volume}
  {65}},\ \bibinfo {pages} {2158} (\bibinfo {year} {1996})}\BibitemShut
  {NoStop}%
\bibitem [{\citenamefont {Kato}(2004)}]{Kato2004}%
  \BibitemOpen
  \bibfield  {author} {\bibinfo {author} {\bibfnamefont {R.}~\bibnamefont
  {Kato}},\ }\href {\doibase 10.1021/cr030655t} {\bibfield  {journal} {\bibinfo
   {journal} {Chem. Rev.}\ }\textbf {\bibinfo {volume} {104}},\ \bibinfo
  {pages} {5319} (\bibinfo {year} {2004})}\BibitemShut {NoStop}%
\bibitem [{\citenamefont {Shimizu}\ \emph {et~al.}(2003)\citenamefont
  {Shimizu}, \citenamefont {Miyagawa}, \citenamefont {Kanoda}, \citenamefont
  {Maesato},\ and\ \citenamefont {Saito}}]{Shimizu2003}%
  \BibitemOpen
  \bibfield  {author} {\bibinfo {author} {\bibfnamefont {Y.}~\bibnamefont
  {Shimizu}}, \bibinfo {author} {\bibfnamefont {K.}~\bibnamefont {Miyagawa}},
  \bibinfo {author} {\bibfnamefont {K.}~\bibnamefont {Kanoda}}, \bibinfo
  {author} {\bibfnamefont {M.}~\bibnamefont {Maesato}}, \ and\ \bibinfo
  {author} {\bibfnamefont {G.}~\bibnamefont {Saito}},\ }\href
  {https://link.aps.org/doi/10.1103/PhysRevLett.91.107001} {\bibfield
  {journal} {\bibinfo  {journal} {Phys. Rev. Lett.}\ }\textbf {\bibinfo
  {volume} {91}},\ \bibinfo {pages} {107001} (\bibinfo {year}
  {2003})}\BibitemShut {NoStop}%
\bibitem [{\citenamefont {Itou}\ \emph {et~al.}(2008)\citenamefont {Itou},
  \citenamefont {Oyamada}, \citenamefont {Maegawa}, \citenamefont {Tamura},\
  and\ \citenamefont {Kato}}]{Itou2008}%
  \BibitemOpen
  \bibfield  {author} {\bibinfo {author} {\bibfnamefont {T.}~\bibnamefont
  {Itou}}, \bibinfo {author} {\bibfnamefont {A.}~\bibnamefont {Oyamada}},
  \bibinfo {author} {\bibfnamefont {S.}~\bibnamefont {Maegawa}}, \bibinfo
  {author} {\bibfnamefont {M.}~\bibnamefont {Tamura}}, \ and\ \bibinfo {author}
  {\bibfnamefont {R.}~\bibnamefont {Kato}},\ }\href
  {https://link.aps.org/doi/10.1103/PhysRevB.77.104413} {\bibfield  {journal}
  {\bibinfo  {journal} {Phys. Rev. B}\ }\textbf {\bibinfo {volume} {77}},\
  \bibinfo {pages} {104413} (\bibinfo {year} {2008})}\BibitemShut {NoStop}%
\bibitem [{\citenamefont {Yamashita}\ \emph {et~al.}(2008)\citenamefont
  {Yamashita}, \citenamefont {Nakazawa}, \citenamefont {Oguni}, \citenamefont
  {Oshima}, \citenamefont {Nojiri}, \citenamefont {Shimizu}, \citenamefont
  {Miyagawa},\ and\ \citenamefont {Kanoda}}]{Yamashita2008}%
  \BibitemOpen
  \bibfield  {author} {\bibinfo {author} {\bibfnamefont {S.}~\bibnamefont
  {Yamashita}}, \bibinfo {author} {\bibfnamefont {Y.}~\bibnamefont {Nakazawa}},
  \bibinfo {author} {\bibfnamefont {M.}~\bibnamefont {Oguni}}, \bibinfo
  {author} {\bibfnamefont {Y.}~\bibnamefont {Oshima}}, \bibinfo {author}
  {\bibfnamefont {H.}~\bibnamefont {Nojiri}}, \bibinfo {author} {\bibfnamefont
  {Y.}~\bibnamefont {Shimizu}}, \bibinfo {author} {\bibfnamefont
  {K.}~\bibnamefont {Miyagawa}}, \ and\ \bibinfo {author} {\bibfnamefont
  {K.}~\bibnamefont {Kanoda}},\ }\href {http://dx.doi.org/10.1038/nphys942}
  {\bibfield  {journal} {\bibinfo  {journal} {Nat. Phys.}\ }\textbf {\bibinfo
  {volume} {4}},\ \bibinfo {pages} {459} (\bibinfo {year} {2008})}\BibitemShut
  {NoStop}%
\bibitem [{\citenamefont {Yamashita}\ \emph {et~al.}(2011)\citenamefont
  {Yamashita}, \citenamefont {Yamamoto}, \citenamefont {Nakazawa},
  \citenamefont {Tamura},\ and\ \citenamefont {Kato}}]{Yamashita2011}%
  \BibitemOpen
  \bibfield  {author} {\bibinfo {author} {\bibfnamefont {S.}~\bibnamefont
  {Yamashita}}, \bibinfo {author} {\bibfnamefont {T.}~\bibnamefont {Yamamoto}},
  \bibinfo {author} {\bibfnamefont {Y.}~\bibnamefont {Nakazawa}}, \bibinfo
  {author} {\bibfnamefont {M.}~\bibnamefont {Tamura}}, \ and\ \bibinfo {author}
  {\bibfnamefont {R.}~\bibnamefont {Kato}},\ }\href
  {http://dx.doi.org/10.1038/ncomms1274 http://10.0.4.14/ncomms1274} {\bibfield
   {journal} {\bibinfo  {journal} {Nat. Commun.}\ }\textbf {\bibinfo {volume}
  {2}},\ \bibinfo {pages} {275} (\bibinfo {year} {2011})}\BibitemShut {NoStop}%
\bibitem [{\citenamefont {Watanabe}\ \emph {et~al.}(2012)\citenamefont
  {Watanabe}, \citenamefont {Yamashita}, \citenamefont {Tonegawa},
  \citenamefont {Oshima}, \citenamefont {Yamamoto}, \citenamefont {Kato},
  \citenamefont {Sheikin}, \citenamefont {Behnia}, \citenamefont {Terashima},
  \citenamefont {Uji}, \citenamefont {Shibauchi},\ and\ \citenamefont
  {Matsuda}}]{Watanabe2012}%
  \BibitemOpen
  \bibfield  {author} {\bibinfo {author} {\bibfnamefont {D.}~\bibnamefont
  {Watanabe}}, \bibinfo {author} {\bibfnamefont {M.}~\bibnamefont {Yamashita}},
  \bibinfo {author} {\bibfnamefont {S.}~\bibnamefont {Tonegawa}}, \bibinfo
  {author} {\bibfnamefont {Y.}~\bibnamefont {Oshima}}, \bibinfo {author}
  {\bibfnamefont {H.~M.}\ \bibnamefont {Yamamoto}}, \bibinfo {author}
  {\bibfnamefont {R.}~\bibnamefont {Kato}}, \bibinfo {author} {\bibfnamefont
  {I.}~\bibnamefont {Sheikin}}, \bibinfo {author} {\bibfnamefont
  {K.}~\bibnamefont {Behnia}}, \bibinfo {author} {\bibfnamefont
  {T.}~\bibnamefont {Terashima}}, \bibinfo {author} {\bibfnamefont
  {S.}~\bibnamefont {Uji}}, \bibinfo {author} {\bibfnamefont {T.}~\bibnamefont
  {Shibauchi}}, \ and\ \bibinfo {author} {\bibfnamefont {Y.}~\bibnamefont
  {Matsuda}},\ }\href {https://doi.org/10.1038/ncomms2082
  http://10.0.4.14/ncomms2082} {\bibfield  {journal} {\bibinfo  {journal} {Nat.
  Commun.}\ }\textbf {\bibinfo {volume} {3}},\ \bibinfo {pages} {1090}
  (\bibinfo {year} {2012})}\BibitemShut {NoStop}%
\bibitem [{\citenamefont {Shimizu}\ \emph {et~al.}(2016)\citenamefont
  {Shimizu}, \citenamefont {Hiramatsu}, \citenamefont {Maesato}, \citenamefont
  {Otsuka}, \citenamefont {Yamochi}, \citenamefont {Ono}, \citenamefont {Itoh},
  \citenamefont {Yoshida}, \citenamefont {Takigawa}, \citenamefont {Yoshida},\
  and\ \citenamefont {Saito}}]{Shimizu2016}%
  \BibitemOpen
  \bibfield  {author} {\bibinfo {author} {\bibfnamefont {Y.}~\bibnamefont
  {Shimizu}}, \bibinfo {author} {\bibfnamefont {T.}~\bibnamefont {Hiramatsu}},
  \bibinfo {author} {\bibfnamefont {M.}~\bibnamefont {Maesato}}, \bibinfo
  {author} {\bibfnamefont {A.}~\bibnamefont {Otsuka}}, \bibinfo {author}
  {\bibfnamefont {H.}~\bibnamefont {Yamochi}}, \bibinfo {author} {\bibfnamefont
  {A.}~\bibnamefont {Ono}}, \bibinfo {author} {\bibfnamefont {M.}~\bibnamefont
  {Itoh}}, \bibinfo {author} {\bibfnamefont {M.}~\bibnamefont {Yoshida}},
  \bibinfo {author} {\bibfnamefont {M.}~\bibnamefont {Takigawa}}, \bibinfo
  {author} {\bibfnamefont {Y.}~\bibnamefont {Yoshida}}, \ and\ \bibinfo
  {author} {\bibfnamefont {G.}~\bibnamefont {Saito}},\ }\href
  {https://link.aps.org/doi/10.1103/PhysRevLett.117.107203} {\bibfield
  {journal} {\bibinfo  {journal} {Phys. Rev. Lett.}\ }\textbf {\bibinfo
  {volume} {117}},\ \bibinfo {pages} {107203} (\bibinfo {year}
  {2016})}\BibitemShut {NoStop}%
\bibitem [{\citenamefont {Yamashita}\ \emph {et~al.}(2010)\citenamefont
  {Yamashita}, \citenamefont {Nakata}, \citenamefont {Senshu}, \citenamefont
  {Nagata}, \citenamefont {Yamamoto}, \citenamefont {Kato}, \citenamefont
  {Shibauchi},\ and\ \citenamefont {Matsuda}}]{Yamashita2010}%
  \BibitemOpen
  \bibfield  {author} {\bibinfo {author} {\bibfnamefont {M.}~\bibnamefont
  {Yamashita}}, \bibinfo {author} {\bibfnamefont {N.}~\bibnamefont {Nakata}},
  \bibinfo {author} {\bibfnamefont {Y.}~\bibnamefont {Senshu}}, \bibinfo
  {author} {\bibfnamefont {M.}~\bibnamefont {Nagata}}, \bibinfo {author}
  {\bibfnamefont {H.~M.}\ \bibnamefont {Yamamoto}}, \bibinfo {author}
  {\bibfnamefont {R.}~\bibnamefont {Kato}}, \bibinfo {author} {\bibfnamefont
  {T.}~\bibnamefont {Shibauchi}}, \ and\ \bibinfo {author} {\bibfnamefont
  {Y.}~\bibnamefont {Matsuda}},\ }\href
  {http://science.sciencemag.org/content/328/5983/1246.abstract} {\bibfield
  {journal} {\bibinfo  {journal} {Science}\ }\textbf {\bibinfo {volume}
  {328}},\ \bibinfo {pages} {1246 LP } (\bibinfo {year} {2010})}\BibitemShut
  {NoStop}%
\bibitem [{\citenamefont {Dressel}\ and\ \citenamefont
  {Pustogow}(2018)}]{Dressel2018}%
  \BibitemOpen
  \bibfield  {author} {\bibinfo {author} {\bibfnamefont {M.}~\bibnamefont
  {Dressel}}\ and\ \bibinfo {author} {\bibfnamefont {A.}~\bibnamefont
  {Pustogow}},\ }\href {http://stacks.iop.org/0953-8984/30/i=20/a=203001}
  {\bibfield  {journal} {\bibinfo  {journal} {J. Phys. Condens. Matter}\
  }\textbf {\bibinfo {volume} {30}},\ \bibinfo {pages} {203001} (\bibinfo
  {year} {2018})}\BibitemShut {NoStop}%
\bibitem [{\citenamefont {Pustogow}\ \emph
  {et~al.}(2018{\natexlab{a}})\citenamefont {Pustogow}, \citenamefont {Saito},
  \citenamefont {Zhukova}, \citenamefont {Gorshunov}, \citenamefont {Kato},
  \citenamefont {Lee}, \citenamefont {Fratini}, \citenamefont
  {Dobrosavljevi{\'{c}}},\ and\ \citenamefont {Dressel}}]{Pustogow2018spinons}%
  \BibitemOpen
  \bibfield  {author} {\bibinfo {author} {\bibfnamefont {A.}~\bibnamefont
  {Pustogow}}, \bibinfo {author} {\bibfnamefont {Y.}~\bibnamefont {Saito}},
  \bibinfo {author} {\bibfnamefont {E.}~\bibnamefont {Zhukova}}, \bibinfo
  {author} {\bibfnamefont {B.}~\bibnamefont {Gorshunov}}, \bibinfo {author}
  {\bibfnamefont {R.}~\bibnamefont {Kato}}, \bibinfo {author} {\bibfnamefont
  {T.-H.}\ \bibnamefont {Lee}}, \bibinfo {author} {\bibfnamefont
  {S.}~\bibnamefont {Fratini}}, \bibinfo {author} {\bibfnamefont
  {V.}~\bibnamefont {Dobrosavljevi{\'{c}}}}, \ and\ \bibinfo {author}
  {\bibfnamefont {M.}~\bibnamefont {Dressel}},\ }\href {\doibase
  10.1103/PhysRevLett.121.056402} {\bibfield  {journal} {\bibinfo  {journal}
  {Phys. Rev. Lett.}\ }\textbf {\bibinfo {volume} {121}},\ \bibinfo {pages}
  {056402} (\bibinfo {year} {2018}{\natexlab{a}})}\BibitemShut {NoStop}%
\bibitem [{Note1()}]{Note1}%
  \BibitemOpen
  \bibinfo {note} {The results from Ref.~\cite {Yamashita2010} have been
  recently challenged in Ref.~\cite {Bourgeois-Hope2019}.}\BibitemShut {Stop}%
\bibitem [{\citenamefont {Pustogow}\ \emph
  {et~al.}(2018{\natexlab{b}})\citenamefont {Pustogow}, \citenamefont {Bories},
  \citenamefont {L{\"{o}}hle}, \citenamefont {R{\"{o}}sslhuber}, \citenamefont
  {Zhukova}, \citenamefont {Gorshunov}, \citenamefont {Tomi{\'{c}}},
  \citenamefont {Schlueter}, \citenamefont {H{\"{u}}bner}, \citenamefont
  {Hiramatsu}, \citenamefont {Yoshida}, \citenamefont {Saito}, \citenamefont
  {Kato}, \citenamefont {Lee}, \citenamefont {Dobrosavljevi{\'{c}}},
  \citenamefont {Fratini},\ and\ \citenamefont {Dressel}}]{Pustogow2018}%
  \BibitemOpen
  \bibfield  {author} {\bibinfo {author} {\bibfnamefont {A.}~\bibnamefont
  {Pustogow}}, \bibinfo {author} {\bibfnamefont {M.}~\bibnamefont {Bories}},
  \bibinfo {author} {\bibfnamefont {A.}~\bibnamefont {L{\"{o}}hle}}, \bibinfo
  {author} {\bibfnamefont {R.}~\bibnamefont {R{\"{o}}sslhuber}}, \bibinfo
  {author} {\bibfnamefont {E.}~\bibnamefont {Zhukova}}, \bibinfo {author}
  {\bibfnamefont {B.}~\bibnamefont {Gorshunov}}, \bibinfo {author}
  {\bibfnamefont {S.}~\bibnamefont {Tomi{\'{c}}}}, \bibinfo {author}
  {\bibfnamefont {J.~A.}\ \bibnamefont {Schlueter}}, \bibinfo {author}
  {\bibfnamefont {R.}~\bibnamefont {H{\"{u}}bner}}, \bibinfo {author}
  {\bibfnamefont {T.}~\bibnamefont {Hiramatsu}}, \bibinfo {author}
  {\bibfnamefont {Y.}~\bibnamefont {Yoshida}}, \bibinfo {author} {\bibfnamefont
  {G.}~\bibnamefont {Saito}}, \bibinfo {author} {\bibfnamefont
  {R.}~\bibnamefont {Kato}}, \bibinfo {author} {\bibfnamefont {T.-H.}\
  \bibnamefont {Lee}}, \bibinfo {author} {\bibfnamefont {V.}~\bibnamefont
  {Dobrosavljevi{\'{c}}}}, \bibinfo {author} {\bibfnamefont {S.}~\bibnamefont
  {Fratini}}, \ and\ \bibinfo {author} {\bibfnamefont {M.}~\bibnamefont
  {Dressel}},\ }\href {\doibase 10.1038/s41563-018-0140-3} {\bibfield
  {journal} {\bibinfo  {journal} {Nat. Mater.}\ }\textbf {\bibinfo {volume}
  {17}},\ \bibinfo {pages} {773} (\bibinfo {year}
  {2018}{\natexlab{b}})}\BibitemShut {NoStop}%
\bibitem [{\citenamefont {Miyagawa}\ \emph {et~al.}(1995)\citenamefont
  {Miyagawa}, \citenamefont {Kawamoto}, \citenamefont {Nakazawa},\ and\
  \citenamefont {Kanoda}}]{Miyagawa1995}%
  \BibitemOpen
  \bibfield  {author} {\bibinfo {author} {\bibfnamefont {K.}~\bibnamefont
  {Miyagawa}}, \bibinfo {author} {\bibfnamefont {A.}~\bibnamefont {Kawamoto}},
  \bibinfo {author} {\bibfnamefont {Y.}~\bibnamefont {Nakazawa}}, \ and\
  \bibinfo {author} {\bibfnamefont {K.}~\bibnamefont {Kanoda}},\ }\href
  {\doibase 10.1103/PhysRevLett.75.1174} {\bibfield  {journal} {\bibinfo
  {journal} {Phys. Rev. Lett.}\ }\textbf {\bibinfo {volume} {75}},\ \bibinfo
  {pages} {1174} (\bibinfo {year} {1995})}\BibitemShut {NoStop}%
\bibitem [{\citenamefont {Shimizu}\ \emph {et~al.}(2006)\citenamefont
  {Shimizu}, \citenamefont {Miyagawa}, \citenamefont {Kanoda}, \citenamefont
  {Maesato},\ and\ \citenamefont {Saito}}]{Shimizu2006}%
  \BibitemOpen
  \bibfield  {author} {\bibinfo {author} {\bibfnamefont {Y.}~\bibnamefont
  {Shimizu}}, \bibinfo {author} {\bibfnamefont {K.}~\bibnamefont {Miyagawa}},
  \bibinfo {author} {\bibfnamefont {K.}~\bibnamefont {Kanoda}}, \bibinfo
  {author} {\bibfnamefont {M.}~\bibnamefont {Maesato}}, \ and\ \bibinfo
  {author} {\bibfnamefont {G.}~\bibnamefont {Saito}},\ }\href
  {https://link.aps.org/doi/10.1103/PhysRevB.73.140407} {\bibfield  {journal}
  {\bibinfo  {journal} {Phys. Rev. B}\ }\textbf {\bibinfo {volume} {73}},\
  \bibinfo {pages} {140407} (\bibinfo {year} {2006})}\BibitemShut {NoStop}%
\bibitem [{\citenamefont {Kandpal}\ \emph {et~al.}(2009)\citenamefont
  {Kandpal}, \citenamefont {Opahle}, \citenamefont {Zhang}, \citenamefont
  {Jeschke},\ and\ \citenamefont {Valent{\'{i}}}}]{Kandpal2009}%
  \BibitemOpen
  \bibfield  {author} {\bibinfo {author} {\bibfnamefont {H.~C.}\ \bibnamefont
  {Kandpal}}, \bibinfo {author} {\bibfnamefont {I.}~\bibnamefont {Opahle}},
  \bibinfo {author} {\bibfnamefont {Y.-Z.}\ \bibnamefont {Zhang}}, \bibinfo
  {author} {\bibfnamefont {H.~O.}\ \bibnamefont {Jeschke}}, \ and\ \bibinfo
  {author} {\bibfnamefont {R.}~\bibnamefont {Valent{\'{i}}}},\ }\href
  {https://link.aps.org/doi/10.1103/PhysRevLett.103.067004} {\bibfield
  {journal} {\bibinfo  {journal} {Phys. Rev. Lett.}\ }\textbf {\bibinfo
  {volume} {103}},\ \bibinfo {pages} {67004} (\bibinfo {year}
  {2009})}\BibitemShut {NoStop}%
\bibitem [{\citenamefont {Nakamura}\ \emph {et~al.}(2009)\citenamefont
  {Nakamura}, \citenamefont {Yoshimoto}, \citenamefont {Kosugi}, \citenamefont
  {Arita},\ and\ \citenamefont {Imada}}]{Nakamura2009}%
  \BibitemOpen
  \bibfield  {author} {\bibinfo {author} {\bibfnamefont {K.}~\bibnamefont
  {Nakamura}}, \bibinfo {author} {\bibfnamefont {Y.}~\bibnamefont {Yoshimoto}},
  \bibinfo {author} {\bibfnamefont {T.}~\bibnamefont {Kosugi}}, \bibinfo
  {author} {\bibfnamefont {R.}~\bibnamefont {Arita}}, \ and\ \bibinfo {author}
  {\bibfnamefont {M.}~\bibnamefont {Imada}},\ }\href {\doibase
  10.1143/JPSJ.78.083710} {\bibfield  {journal} {\bibinfo  {journal} {J. Phys.
  Soc. Jpn.}\ }\textbf {\bibinfo {volume} {78}},\ \bibinfo {pages} {83710}
  (\bibinfo {year} {2009})}\BibitemShut {NoStop}%
\bibitem [{\citenamefont {Lunkenheimer}\ \emph {et~al.}(2012)\citenamefont
  {Lunkenheimer}, \citenamefont {M{\"{u}}ller}, \citenamefont {Krohns},
  \citenamefont {Schrettle}, \citenamefont {Loidl}, \citenamefont {Hartmann},
  \citenamefont {Rommel}, \citenamefont {de~Souza}, \citenamefont {Hotta},
  \citenamefont {Schlueter},\ and\ \citenamefont {Lang}}]{Lunkenheimer2012}%
  \BibitemOpen
  \bibfield  {author} {\bibinfo {author} {\bibfnamefont {P.}~\bibnamefont
  {Lunkenheimer}}, \bibinfo {author} {\bibfnamefont {J.}~\bibnamefont
  {M{\"{u}}ller}}, \bibinfo {author} {\bibfnamefont {S.}~\bibnamefont
  {Krohns}}, \bibinfo {author} {\bibfnamefont {F.}~\bibnamefont {Schrettle}},
  \bibinfo {author} {\bibfnamefont {A.}~\bibnamefont {Loidl}}, \bibinfo
  {author} {\bibfnamefont {B.}~\bibnamefont {Hartmann}}, \bibinfo {author}
  {\bibfnamefont {R.}~\bibnamefont {Rommel}}, \bibinfo {author} {\bibfnamefont
  {M.}~\bibnamefont {de~Souza}}, \bibinfo {author} {\bibfnamefont
  {C.}~\bibnamefont {Hotta}}, \bibinfo {author} {\bibfnamefont {J.~A.}\
  \bibnamefont {Schlueter}}, \ and\ \bibinfo {author} {\bibfnamefont
  {M.}~\bibnamefont {Lang}},\ }\href {http://dx.doi.org/10.1038/nmat3400
  http://10.0.4.14/nmat3400
  https://www.nature.com/articles/nmat3400{\#}supplementary-information}
  {\bibfield  {journal} {\bibinfo  {journal} {Nat. Mater.}\ }\textbf {\bibinfo
  {volume} {11}},\ \bibinfo {pages} {755} (\bibinfo {year} {2012})}\BibitemShut
  {NoStop}%
\bibitem [{\citenamefont {Matsuura}\ \emph {et~al.}(2019)\citenamefont
  {Matsuura}, \citenamefont {Sasaki}, \citenamefont {Iguchi}, \citenamefont
  {Gati}, \citenamefont {M{\"{u}}ller}, \citenamefont {Stockert}, \citenamefont
  {Piovano}, \citenamefont {B{\"{o}}hm}, \citenamefont {Park}, \citenamefont
  {Biswas}, \citenamefont {Winter}, \citenamefont {Valent{\'{i}}},
  \citenamefont {Nakao},\ and\ \citenamefont {Lang}}]{Matsuura2019}%
  \BibitemOpen
  \bibfield  {author} {\bibinfo {author} {\bibfnamefont {M.}~\bibnamefont
  {Matsuura}}, \bibinfo {author} {\bibfnamefont {T.}~\bibnamefont {Sasaki}},
  \bibinfo {author} {\bibfnamefont {S.}~\bibnamefont {Iguchi}}, \bibinfo
  {author} {\bibfnamefont {E.}~\bibnamefont {Gati}}, \bibinfo {author}
  {\bibfnamefont {J.}~\bibnamefont {M{\"{u}}ller}}, \bibinfo {author}
  {\bibfnamefont {O.}~\bibnamefont {Stockert}}, \bibinfo {author}
  {\bibfnamefont {A.}~\bibnamefont {Piovano}}, \bibinfo {author} {\bibfnamefont
  {M.}~\bibnamefont {B{\"{o}}hm}}, \bibinfo {author} {\bibfnamefont {J.~T.}\
  \bibnamefont {Park}}, \bibinfo {author} {\bibfnamefont {S.}~\bibnamefont
  {Biswas}}, \bibinfo {author} {\bibfnamefont {S.~M.}\ \bibnamefont {Winter}},
  \bibinfo {author} {\bibfnamefont {R.}~\bibnamefont {Valent{\'{i}}}}, \bibinfo
  {author} {\bibfnamefont {A.}~\bibnamefont {Nakao}}, \ and\ \bibinfo {author}
  {\bibfnamefont {M.}~\bibnamefont {Lang}},\ }\href {\doibase
  10.1103/PhysRevLett.123.027601} {\bibfield  {journal} {\bibinfo  {journal}
  {Phys. Rev. Lett.}\ }\textbf {\bibinfo {volume} {123}},\ \bibinfo {pages}
  {27601} (\bibinfo {year} {2019})}\BibitemShut {NoStop}%
\bibitem [{\citenamefont {Guterding}\ \emph {et~al.}(2015)\citenamefont
  {Guterding}, \citenamefont {Valent{\'{i}}},\ and\ \citenamefont
  {Jeschke}}]{Guterding2015}%
  \BibitemOpen
  \bibfield  {author} {\bibinfo {author} {\bibfnamefont {D.}~\bibnamefont
  {Guterding}}, \bibinfo {author} {\bibfnamefont {R.}~\bibnamefont
  {Valent{\'{i}}}}, \ and\ \bibinfo {author} {\bibfnamefont {H.~O.}\
  \bibnamefont {Jeschke}},\ }\href {\doibase 10.1103/PhysRevB.92.081109}
  {\bibfield  {journal} {\bibinfo  {journal} {Phys. Rev. B}\ }\textbf {\bibinfo
  {volume} {92}},\ \bibinfo {pages} {81109} (\bibinfo {year}
  {2015})}\BibitemShut {NoStop}%
\bibitem [{\citenamefont {Huse}\ and\ \citenamefont {Elser}(1988)}]{Huse1988}%
  \BibitemOpen
  \bibfield  {author} {\bibinfo {author} {\bibfnamefont {D.~A.}\ \bibnamefont
  {Huse}}\ and\ \bibinfo {author} {\bibfnamefont {V.}~\bibnamefont {Elser}},\
  }\href {\doibase 10.1103/PhysRevLett.60.2531} {\bibfield  {journal} {\bibinfo
   {journal} {Phys. Rev. Lett.}\ }\textbf {\bibinfo {volume} {60}},\ \bibinfo
  {pages} {2531} (\bibinfo {year} {1988})}\BibitemShut {NoStop}%
\bibitem [{\citenamefont {Furukawa}\ \emph {et~al.}(2015)\citenamefont
  {Furukawa}, \citenamefont {Miyagawa}, \citenamefont {Itou}, \citenamefont
  {Ito}, \citenamefont {Taniguchi}, \citenamefont {Saito}, \citenamefont
  {Iguchi}, \citenamefont {Sasaki},\ and\ \citenamefont
  {Kanoda}}]{Furukawa2015b}%
  \BibitemOpen
  \bibfield  {author} {\bibinfo {author} {\bibfnamefont {T.}~\bibnamefont
  {Furukawa}}, \bibinfo {author} {\bibfnamefont {K.}~\bibnamefont {Miyagawa}},
  \bibinfo {author} {\bibfnamefont {T.}~\bibnamefont {Itou}}, \bibinfo {author}
  {\bibfnamefont {M.}~\bibnamefont {Ito}}, \bibinfo {author} {\bibfnamefont
  {H.}~\bibnamefont {Taniguchi}}, \bibinfo {author} {\bibfnamefont
  {M.}~\bibnamefont {Saito}}, \bibinfo {author} {\bibfnamefont
  {S.}~\bibnamefont {Iguchi}}, \bibinfo {author} {\bibfnamefont
  {T.}~\bibnamefont {Sasaki}}, \ and\ \bibinfo {author} {\bibfnamefont
  {K.}~\bibnamefont {Kanoda}},\ }\href
  {https://link.aps.org/doi/10.1103/PhysRevLett.115.077001} {\bibfield
  {journal} {\bibinfo  {journal} {Phys. Rev. Lett.}\ }\textbf {\bibinfo
  {volume} {115}},\ \bibinfo {pages} {77001} (\bibinfo {year}
  {2015})}\BibitemShut {NoStop}%
\bibitem [{\citenamefont {Dressel}\ \emph {et~al.}(2016)\citenamefont
  {Dressel}, \citenamefont {Lazi{\'{c}}}, \citenamefont {Pustogow},
  \citenamefont {Zhukova}, \citenamefont {Gorshunov}, \citenamefont
  {Schlueter}, \citenamefont {Milat}, \citenamefont {Gumhalter},\ and\
  \citenamefont {Tomi{\'{c}}}}]{Dressel2016}%
  \BibitemOpen
  \bibfield  {author} {\bibinfo {author} {\bibfnamefont {M.}~\bibnamefont
  {Dressel}}, \bibinfo {author} {\bibfnamefont {P.}~\bibnamefont
  {Lazi{\'{c}}}}, \bibinfo {author} {\bibfnamefont {A.}~\bibnamefont
  {Pustogow}}, \bibinfo {author} {\bibfnamefont {E.}~\bibnamefont {Zhukova}},
  \bibinfo {author} {\bibfnamefont {B.}~\bibnamefont {Gorshunov}}, \bibinfo
  {author} {\bibfnamefont {J.~A.}\ \bibnamefont {Schlueter}}, \bibinfo {author}
  {\bibfnamefont {O.}~\bibnamefont {Milat}}, \bibinfo {author} {\bibfnamefont
  {B.}~\bibnamefont {Gumhalter}}, \ and\ \bibinfo {author} {\bibfnamefont
  {S.}~\bibnamefont {Tomi{\'{c}}}},\ }\href
  {https://link.aps.org/doi/10.1103/PhysRevB.93.081201} {\bibfield  {journal}
  {\bibinfo  {journal} {Phys. Rev. B}\ }\textbf {\bibinfo {volume} {93}},\
  \bibinfo {pages} {81201} (\bibinfo {year} {2016})}\BibitemShut {NoStop}%
\bibitem [{\citenamefont {Pinteri{\'{c}}}\ \emph {et~al.}(2018)\citenamefont
  {Pinteri{\'{c}}}, \citenamefont {{Rivas G{\'{o}}ngora}}, \citenamefont
  {Rapljenovi{\'{c}}}, \citenamefont {Ivek}, \citenamefont {{\v{C}}ulo},
  \citenamefont {Korin-Hamzi{\'{c}}}, \citenamefont {Milat}, \citenamefont
  {Gumhalter}, \citenamefont {Lazi{\'{c}}}, \citenamefont {{Sanz Alonso}},
  \citenamefont {Li}, \citenamefont {Pustogow}, \citenamefont {{Gorgen
  Lesseux}}, \citenamefont {Dressel},\ and\ \citenamefont
  {Tomi{\'{c}}}}]{Pinteric2018}%
  \BibitemOpen
  \bibfield  {author} {\bibinfo {author} {\bibfnamefont {M.}~\bibnamefont
  {Pinteri{\'{c}}}}, \bibinfo {author} {\bibfnamefont {D.}~\bibnamefont {{Rivas
  G{\'{o}}ngora}}}, \bibinfo {author} {\bibfnamefont {{\v{Z}}.}~\bibnamefont
  {Rapljenovi{\'{c}}}}, \bibinfo {author} {\bibfnamefont {T.}~\bibnamefont
  {Ivek}}, \bibinfo {author} {\bibfnamefont {M.}~\bibnamefont {{\v{C}}ulo}},
  \bibinfo {author} {\bibfnamefont {B.}~\bibnamefont {Korin-Hamzi{\'{c}}}},
  \bibinfo {author} {\bibfnamefont {O.}~\bibnamefont {Milat}}, \bibinfo
  {author} {\bibfnamefont {B.}~\bibnamefont {Gumhalter}}, \bibinfo {author}
  {\bibfnamefont {P.}~\bibnamefont {Lazi{\'{c}}}}, \bibinfo {author}
  {\bibfnamefont {M.}~\bibnamefont {{Sanz Alonso}}}, \bibinfo {author}
  {\bibfnamefont {W.}~\bibnamefont {Li}}, \bibinfo {author} {\bibfnamefont
  {A.}~\bibnamefont {Pustogow}}, \bibinfo {author} {\bibfnamefont
  {G.}~\bibnamefont {{Gorgen Lesseux}}}, \bibinfo {author} {\bibfnamefont
  {M.}~\bibnamefont {Dressel}}, \ and\ \bibinfo {author} {\bibfnamefont
  {S.}~\bibnamefont {Tomi{\'{c}}}},\ }\href {\doibase 10.3390/cryst8050190}
  {\bibfield  {journal} {\bibinfo  {journal} {Crystals}\ }\textbf {\bibinfo
  {volume} {8}},\ \bibinfo {pages} {190} (\bibinfo {year} {2018})}\BibitemShut
  {NoStop}%
\bibitem [{\citenamefont {Itou}\ \emph {et~al.}(2017)\citenamefont {Itou},
  \citenamefont {Watanabe}, \citenamefont {Maegawa}, \citenamefont {Tajima},
  \citenamefont {Tajima}, \citenamefont {Kubo}, \citenamefont {Kato},\ and\
  \citenamefont {Kanoda}}]{Itou2017}%
  \BibitemOpen
  \bibfield  {author} {\bibinfo {author} {\bibfnamefont {T.}~\bibnamefont
  {Itou}}, \bibinfo {author} {\bibfnamefont {E.}~\bibnamefont {Watanabe}},
  \bibinfo {author} {\bibfnamefont {S.}~\bibnamefont {Maegawa}}, \bibinfo
  {author} {\bibfnamefont {A.}~\bibnamefont {Tajima}}, \bibinfo {author}
  {\bibfnamefont {N.}~\bibnamefont {Tajima}}, \bibinfo {author} {\bibfnamefont
  {K.}~\bibnamefont {Kubo}}, \bibinfo {author} {\bibfnamefont {R.}~\bibnamefont
  {Kato}}, \ and\ \bibinfo {author} {\bibfnamefont {K.}~\bibnamefont
  {Kanoda}},\ }\href
  {http://advances.sciencemag.org/content/3/8/e1601594.abstract} {\bibfield
  {journal} {\bibinfo  {journal} {Sci. Adv.}\ }\textbf {\bibinfo {volume}
  {3}},\ \bibinfo {pages} {e1601594} (\bibinfo {year} {2017})}\BibitemShut
  {NoStop}%
\bibitem [{\citenamefont {Lazi{\'{c}}}\ \emph {et~al.}(2018)\citenamefont
  {Lazi{\'{c}}}, \citenamefont {Pinteri{\'{c}}}, \citenamefont {{Rivas
  G{\'{o}}ngora}}, \citenamefont {Pustogow}, \citenamefont {Treptow},
  \citenamefont {Ivek}, \citenamefont {Milat}, \citenamefont {Gumhalter},
  \citenamefont {Do{\v{s}}li{\'{c}}}, \citenamefont {Dressel},\ and\
  \citenamefont {Tomi{\'{c}}}}]{Lazic2018}%
  \BibitemOpen
  \bibfield  {author} {\bibinfo {author} {\bibfnamefont {P.}~\bibnamefont
  {Lazi{\'{c}}}}, \bibinfo {author} {\bibfnamefont {M.}~\bibnamefont
  {Pinteri{\'{c}}}}, \bibinfo {author} {\bibfnamefont {D.}~\bibnamefont {{Rivas
  G{\'{o}}ngora}}}, \bibinfo {author} {\bibfnamefont {A.}~\bibnamefont
  {Pustogow}}, \bibinfo {author} {\bibfnamefont {K.}~\bibnamefont {Treptow}},
  \bibinfo {author} {\bibfnamefont {T.}~\bibnamefont {Ivek}}, \bibinfo {author}
  {\bibfnamefont {O.}~\bibnamefont {Milat}}, \bibinfo {author} {\bibfnamefont
  {B.}~\bibnamefont {Gumhalter}}, \bibinfo {author} {\bibfnamefont
  {N.}~\bibnamefont {Do{\v{s}}li{\'{c}}}}, \bibinfo {author} {\bibfnamefont
  {M.}~\bibnamefont {Dressel}}, \ and\ \bibinfo {author} {\bibfnamefont
  {S.}~\bibnamefont {Tomi{\'{c}}}},\ }\href {\doibase
  10.1103/PhysRevB.97.245134} {\bibfield  {journal} {\bibinfo  {journal} {Phys.
  Rev. B}\ }\textbf {\bibinfo {volume} {97}},\ \bibinfo {pages} {245134}
  (\bibinfo {year} {2018})}\BibitemShut {NoStop}%
\bibitem [{\citenamefont {Riedl}\ \emph {et~al.}(2019)\citenamefont {Riedl},
  \citenamefont {Valent{\'{i}}},\ and\ \citenamefont {Winter}}]{Riedl2019}%
  \BibitemOpen
  \bibfield  {author} {\bibinfo {author} {\bibfnamefont {K.}~\bibnamefont
  {Riedl}}, \bibinfo {author} {\bibfnamefont {R.}~\bibnamefont
  {Valent{\'{i}}}}, \ and\ \bibinfo {author} {\bibfnamefont {S.~M.}\
  \bibnamefont {Winter}},\ }\href {\doibase 10.1038/s41467-019-10604-3}
  {\bibfield  {journal} {\bibinfo  {journal} {Nat. Commun.}\ }\textbf {\bibinfo
  {volume} {10}},\ \bibinfo {pages} {2561} (\bibinfo {year}
  {2019})}\BibitemShut {NoStop}%
\bibitem [{\citenamefont {Powell}\ \emph {et~al.}(2017)\citenamefont {Powell},
  \citenamefont {Kenny},\ and\ \citenamefont {Merino}}]{Powell2017}%
  \BibitemOpen
  \bibfield  {author} {\bibinfo {author} {\bibfnamefont {B.~J.}\ \bibnamefont
  {Powell}}, \bibinfo {author} {\bibfnamefont {E.~P.}\ \bibnamefont {Kenny}}, \
  and\ \bibinfo {author} {\bibfnamefont {J.}~\bibnamefont {Merino}},\ }\href
  {\doibase 10.1103/PhysRevLett.119.087204} {\bibfield  {journal} {\bibinfo
  {journal} {Phys. Rev. Lett.}\ }\textbf {\bibinfo {volume} {119}},\ \bibinfo
  {pages} {87204} (\bibinfo {year} {2017})}\BibitemShut {NoStop}%
\bibitem [{\citenamefont {Motrunich}(2005)}]{Motrunich2005}%
  \BibitemOpen
  \bibfield  {author} {\bibinfo {author} {\bibfnamefont {O.~I.}\ \bibnamefont
  {Motrunich}},\ }\href {\doibase 10.1103/PhysRevB.72.045105} {\bibfield
  {journal} {\bibinfo  {journal} {Phys. Rev. B}\ }\textbf {\bibinfo {volume}
  {72}},\ \bibinfo {pages} {45105} (\bibinfo {year} {2005})}\BibitemShut
  {NoStop}%
\bibitem [{\citenamefont {Yasin}\ \emph {et~al.}(2012)\citenamefont {Yasin},
  \citenamefont {Rose}, \citenamefont {Dumm}, \citenamefont {Drichko},
  \citenamefont {Dressel}, \citenamefont {Schlueter}, \citenamefont
  {Zhilyaeva}, \citenamefont {Torunova},\ and\ \citenamefont
  {Lyubovskaya}}]{Yasin2012}%
  \BibitemOpen
  \bibfield  {author} {\bibinfo {author} {\bibfnamefont {S.}~\bibnamefont
  {Yasin}}, \bibinfo {author} {\bibfnamefont {E.}~\bibnamefont {Rose}},
  \bibinfo {author} {\bibfnamefont {M.}~\bibnamefont {Dumm}}, \bibinfo {author}
  {\bibfnamefont {N.}~\bibnamefont {Drichko}}, \bibinfo {author} {\bibfnamefont
  {M.}~\bibnamefont {Dressel}}, \bibinfo {author} {\bibfnamefont {J.~A.}\
  \bibnamefont {Schlueter}}, \bibinfo {author} {\bibfnamefont {E.~I.}\
  \bibnamefont {Zhilyaeva}}, \bibinfo {author} {\bibfnamefont {S.}~\bibnamefont
  {Torunova}}, \ and\ \bibinfo {author} {\bibfnamefont {R.~N.}\ \bibnamefont
  {Lyubovskaya}},\ }\href {\doibase
  https://doi.org/10.1016/j.physb.2012.01.007} {\bibfield  {journal} {\bibinfo
  {journal} {Physica B: Condens. Matter}\ }\textbf {\bibinfo {volume} {407}},\
  \bibinfo {pages} {1689} (\bibinfo {year} {2012})}\BibitemShut {NoStop}%
\bibitem [{\citenamefont {Drichko}\ \emph {et~al.}(2014)\citenamefont
  {Drichko}, \citenamefont {Beyer}, \citenamefont {Rose}, \citenamefont
  {Dressel}, \citenamefont {Schlueter}, \citenamefont {Turunova}, \citenamefont
  {Zhilyaeva},\ and\ \citenamefont {Lyubovskaya}}]{Drichko2014}%
  \BibitemOpen
  \bibfield  {author} {\bibinfo {author} {\bibfnamefont {N.}~\bibnamefont
  {Drichko}}, \bibinfo {author} {\bibfnamefont {R.}~\bibnamefont {Beyer}},
  \bibinfo {author} {\bibfnamefont {E.}~\bibnamefont {Rose}}, \bibinfo {author}
  {\bibfnamefont {M.}~\bibnamefont {Dressel}}, \bibinfo {author} {\bibfnamefont
  {J.~A.}\ \bibnamefont {Schlueter}}, \bibinfo {author} {\bibfnamefont {S.~A.}\
  \bibnamefont {Turunova}}, \bibinfo {author} {\bibfnamefont {E.~I.}\
  \bibnamefont {Zhilyaeva}}, \ and\ \bibinfo {author} {\bibfnamefont {R.~N.}\
  \bibnamefont {Lyubovskaya}},\ }\href
  {https://link.aps.org/doi/10.1103/PhysRevB.89.075133} {\bibfield  {journal}
  {\bibinfo  {journal} {Phys. Rev. B}\ }\textbf {\bibinfo {volume} {89}},\
  \bibinfo {pages} {75133} (\bibinfo {year} {2014})}\BibitemShut {NoStop}%
\bibitem [{\citenamefont {L{\"{o}}hle}\ \emph {et~al.}(2017)\citenamefont
  {L{\"{o}}hle}, \citenamefont {Rose}, \citenamefont {Singh}, \citenamefont
  {Beyer}, \citenamefont {Tafra}, \citenamefont {R}, \citenamefont {Zhilyaeva},
  \citenamefont {Lyubovskaya},\ and\ \citenamefont {Dressel}}]{Lohle2017}%
  \BibitemOpen
  \bibfield  {author} {\bibinfo {author} {\bibfnamefont {A.}~\bibnamefont
  {L{\"{o}}hle}}, \bibinfo {author} {\bibfnamefont {E.}~\bibnamefont {Rose}},
  \bibinfo {author} {\bibfnamefont {S.}~\bibnamefont {Singh}}, \bibinfo
  {author} {\bibfnamefont {R.}~\bibnamefont {Beyer}}, \bibinfo {author}
  {\bibfnamefont {E.}~\bibnamefont {Tafra}}, \bibinfo {author} {\bibfnamefont
  {I.}~\bibnamefont {R}}, \bibinfo {author} {\bibfnamefont {E.~I.}\
  \bibnamefont {Zhilyaeva}}, \bibinfo {author} {\bibfnamefont {R.~N.}\
  \bibnamefont {Lyubovskaya}}, \ and\ \bibinfo {author} {\bibfnamefont
  {M.}~\bibnamefont {Dressel}},\ }\href
  {http://stacks.iop.org/0953-8984/29/i=5/a=055601} {\bibfield  {journal}
  {\bibinfo  {journal} {J. Phys. Condens. Matter}\ }\textbf {\bibinfo {volume}
  {29}},\ \bibinfo {pages} {55601} (\bibinfo {year} {2017})}\BibitemShut
  {NoStop}%
\bibitem [{\citenamefont {Ivek}\ \emph {et~al.}(2017)\citenamefont {Ivek},
  \citenamefont {Beyer}, \citenamefont {Badalov}, \citenamefont {{\v{C}}ulo},
  \citenamefont {Tomi{\'{c}}}, \citenamefont {Schlueter}, \citenamefont
  {Zhilyaeva}, \citenamefont {Lyubovskaya},\ and\ \citenamefont
  {Dressel}}]{Ivek2017}%
  \BibitemOpen
  \bibfield  {author} {\bibinfo {author} {\bibfnamefont {T.}~\bibnamefont
  {Ivek}}, \bibinfo {author} {\bibfnamefont {R.}~\bibnamefont {Beyer}},
  \bibinfo {author} {\bibfnamefont {S.}~\bibnamefont {Badalov}}, \bibinfo
  {author} {\bibfnamefont {M.}~\bibnamefont {{\v{C}}ulo}}, \bibinfo {author}
  {\bibfnamefont {S.}~\bibnamefont {Tomi{\'{c}}}}, \bibinfo {author}
  {\bibfnamefont {J.~A.}\ \bibnamefont {Schlueter}}, \bibinfo {author}
  {\bibfnamefont {E.~I.}\ \bibnamefont {Zhilyaeva}}, \bibinfo {author}
  {\bibfnamefont {R.~N.}\ \bibnamefont {Lyubovskaya}}, \ and\ \bibinfo {author}
  {\bibfnamefont {M.}~\bibnamefont {Dressel}},\ }\href
  {https://link.aps.org/doi/10.1103/PhysRevB.96.085116} {\bibfield  {journal}
  {\bibinfo  {journal} {Phys. Rev. B}\ }\textbf {\bibinfo {volume} {96}},\
  \bibinfo {pages} {85116} (\bibinfo {year} {2017})}\BibitemShut {NoStop}%
\bibitem [{\citenamefont {Gati}\ \emph
  {et~al.}(2018{\natexlab{a}})\citenamefont {Gati}, \citenamefont {Fischer},
  \citenamefont {Lunkenheimer}, \citenamefont {Zielke}, \citenamefont
  {K{\"{o}}hler}, \citenamefont {Kolb}, \citenamefont {von Nidda},
  \citenamefont {Winter}, \citenamefont {Schubert}, \citenamefont {Schlueter},
  \citenamefont {Jeschke}, \citenamefont {Valent{\'{i}}},\ and\ \citenamefont
  {Lang}}]{Gati2018a}%
  \BibitemOpen
  \bibfield  {author} {\bibinfo {author} {\bibfnamefont {E.}~\bibnamefont
  {Gati}}, \bibinfo {author} {\bibfnamefont {J.~K.~H.}\ \bibnamefont
  {Fischer}}, \bibinfo {author} {\bibfnamefont {P.}~\bibnamefont
  {Lunkenheimer}}, \bibinfo {author} {\bibfnamefont {D.}~\bibnamefont
  {Zielke}}, \bibinfo {author} {\bibfnamefont {S.}~\bibnamefont
  {K{\"{o}}hler}}, \bibinfo {author} {\bibfnamefont {F.}~\bibnamefont {Kolb}},
  \bibinfo {author} {\bibfnamefont {H.-A.~K.}\ \bibnamefont {von Nidda}},
  \bibinfo {author} {\bibfnamefont {S.~M.}\ \bibnamefont {Winter}}, \bibinfo
  {author} {\bibfnamefont {H.}~\bibnamefont {Schubert}}, \bibinfo {author}
  {\bibfnamefont {J.~A.}\ \bibnamefont {Schlueter}}, \bibinfo {author}
  {\bibfnamefont {H.~O.}\ \bibnamefont {Jeschke}}, \bibinfo {author}
  {\bibfnamefont {R.}~\bibnamefont {Valent{\'{i}}}}, \ and\ \bibinfo {author}
  {\bibfnamefont {M.}~\bibnamefont {Lang}},\ }\href {\doibase
  10.1103/PhysRevLett.120.247601} {\bibfield  {journal} {\bibinfo  {journal}
  {Phys. Rev. Lett.}\ }\textbf {\bibinfo {volume} {120}},\ \bibinfo {pages}
  {247601} (\bibinfo {year} {2018}{\natexlab{a}})}\BibitemShut {NoStop}%
\bibitem [{\citenamefont {Hemmida}\ \emph {et~al.}(2018)\citenamefont
  {Hemmida}, \citenamefont {von Nidda}, \citenamefont {Miksch}, \citenamefont
  {Samoilenko}, \citenamefont {Pustogow}, \citenamefont {Widmann},
  \citenamefont {Henderson}, \citenamefont {Siegrist}, \citenamefont
  {Schlueter}, \citenamefont {Loidl},\ and\ \citenamefont
  {Dressel}}]{Hemmida2018}%
  \BibitemOpen
  \bibfield  {author} {\bibinfo {author} {\bibfnamefont {M.}~\bibnamefont
  {Hemmida}}, \bibinfo {author} {\bibfnamefont {H.-A.~K.}\ \bibnamefont {von
  Nidda}}, \bibinfo {author} {\bibfnamefont {B.}~\bibnamefont {Miksch}},
  \bibinfo {author} {\bibfnamefont {L.~L.}\ \bibnamefont {Samoilenko}},
  \bibinfo {author} {\bibfnamefont {A.}~\bibnamefont {Pustogow}}, \bibinfo
  {author} {\bibfnamefont {S.}~\bibnamefont {Widmann}}, \bibinfo {author}
  {\bibfnamefont {A.}~\bibnamefont {Henderson}}, \bibinfo {author}
  {\bibfnamefont {T.}~\bibnamefont {Siegrist}}, \bibinfo {author}
  {\bibfnamefont {J.~A.}\ \bibnamefont {Schlueter}}, \bibinfo {author}
  {\bibfnamefont {A.}~\bibnamefont {Loidl}}, \ and\ \bibinfo {author}
  {\bibfnamefont {M.}~\bibnamefont {Dressel}},\ }\href {\doibase
  10.1103/PhysRevB.98.241202} {\bibfield  {journal} {\bibinfo  {journal} {Phys.
  Rev. B}\ }\textbf {\bibinfo {volume} {98}},\ \bibinfo {pages} {241202(R)}
  (\bibinfo {year} {2018})}\BibitemShut {NoStop}%
\bibitem [{\citenamefont {Gati}\ \emph
  {et~al.}(2018{\natexlab{b}})\citenamefont {Gati}, \citenamefont {Winter},
  \citenamefont {Schlueter}, \citenamefont {Schubert}, \citenamefont
  {M{\"{u}}ller},\ and\ \citenamefont {Lang}}]{Gati2018}%
  \BibitemOpen
  \bibfield  {author} {\bibinfo {author} {\bibfnamefont {E.}~\bibnamefont
  {Gati}}, \bibinfo {author} {\bibfnamefont {S.~M.}\ \bibnamefont {Winter}},
  \bibinfo {author} {\bibfnamefont {J.~A.}\ \bibnamefont {Schlueter}}, \bibinfo
  {author} {\bibfnamefont {H.}~\bibnamefont {Schubert}}, \bibinfo {author}
  {\bibfnamefont {J.}~\bibnamefont {M{\"{u}}ller}}, \ and\ \bibinfo {author}
  {\bibfnamefont {M.}~\bibnamefont {Lang}},\ }\href {\doibase
  10.1103/PhysRevB.97.075115} {\bibfield  {journal} {\bibinfo  {journal} {Phys.
  Rev. B}\ }\textbf {\bibinfo {volume} {97}},\ \bibinfo {pages} {75115}
  (\bibinfo {year} {2018}{\natexlab{b}})}\BibitemShut {NoStop}%
\bibitem [{\citenamefont {Hassan}\ \emph {et~al.}(2018)\citenamefont {Hassan},
  \citenamefont {Cunningham}, \citenamefont {Mourigal}, \citenamefont
  {Zhilyaeva}, \citenamefont {Torunova}, \citenamefont {Lyubovskaya},
  \citenamefont {Schlueter},\ and\ \citenamefont {Drichko}}]{Hassan2018}%
  \BibitemOpen
  \bibfield  {author} {\bibinfo {author} {\bibfnamefont {N.}~\bibnamefont
  {Hassan}}, \bibinfo {author} {\bibfnamefont {S.}~\bibnamefont {Cunningham}},
  \bibinfo {author} {\bibfnamefont {M.}~\bibnamefont {Mourigal}}, \bibinfo
  {author} {\bibfnamefont {E.~I.}\ \bibnamefont {Zhilyaeva}}, \bibinfo {author}
  {\bibfnamefont {S.~A.}\ \bibnamefont {Torunova}}, \bibinfo {author}
  {\bibfnamefont {R.~N.}\ \bibnamefont {Lyubovskaya}}, \bibinfo {author}
  {\bibfnamefont {J.~A.}\ \bibnamefont {Schlueter}}, \ and\ \bibinfo {author}
  {\bibfnamefont {N.}~\bibnamefont {Drichko}},\ }\href
  {http://science.sciencemag.org/content/360/6393/1101.abstract} {\bibfield
  {journal} {\bibinfo  {journal} {Science}\ }\textbf {\bibinfo {volume}
  {360}},\ \bibinfo {pages} {1101 LP } (\bibinfo {year} {2018})}\BibitemShut
  {NoStop}%
\bibitem [{\citenamefont {Hassan}\ \emph {et~al.}(2019)\citenamefont {Hassan},
  \citenamefont {Thirunavukkuarasu}, \citenamefont {Lu}, \citenamefont
  {Smirnov}, \citenamefont {Zhilyaeva}, \citenamefont {Torunova}, \citenamefont
  {Lyubovskaya},\ and\ \citenamefont {Drichko}}]{Hassan2019}%
  \BibitemOpen
  \bibfield  {author} {\bibinfo {author} {\bibfnamefont {N.~M.}\ \bibnamefont
  {Hassan}}, \bibinfo {author} {\bibfnamefont {K.}~\bibnamefont
  {Thirunavukkuarasu}}, \bibinfo {author} {\bibfnamefont {Z.}~\bibnamefont
  {Lu}}, \bibinfo {author} {\bibfnamefont {D.}~\bibnamefont {Smirnov}},
  \bibinfo {author} {\bibfnamefont {E.}~\bibnamefont {Zhilyaeva}}, \bibinfo
  {author} {\bibfnamefont {S.}~\bibnamefont {Torunova}}, \bibinfo {author}
  {\bibfnamefont {R.}~\bibnamefont {Lyubovskaya}}, \ and\ \bibinfo {author}
  {\bibfnamefont {N.}~\bibnamefont {Drichko}},\ }\href@noop {} {\  (\bibinfo
  {year} {2019})},\ \Eprint {http://arxiv.org/abs/1905.12740}
  {arXiv:1905.12740} \BibitemShut {NoStop}%
\bibitem [{Note2()}]{Note2}%
  \BibitemOpen
  \bibinfo {note} {While for $\kappa $-(BEDT-TTF)$_2$Hg(SCN)$_2$Cl\
  $t_d/t'\approx 3$, the paradigmatic $\kappa $-phase materials $\kappa $-CuCN,
  $\kappa $-AgCN\ and $\kappa $-CuCl\ have larger ratios of approximately
  4--5.}\BibitemShut {Stop}%
\bibitem [{\citenamefont {Yue}\ \emph {et~al.}(2010)\citenamefont {Yue},
  \citenamefont {Yamamoto}, \citenamefont {Uruichi}, \citenamefont {Nakano},
  \citenamefont {Yakushi}, \citenamefont {Yamada}, \citenamefont {Hiejima},\
  and\ \citenamefont {Kawamoto}}]{Yue2010}%
  \BibitemOpen
  \bibfield  {author} {\bibinfo {author} {\bibfnamefont {Y.}~\bibnamefont
  {Yue}}, \bibinfo {author} {\bibfnamefont {K.}~\bibnamefont {Yamamoto}},
  \bibinfo {author} {\bibfnamefont {M.}~\bibnamefont {Uruichi}}, \bibinfo
  {author} {\bibfnamefont {C.}~\bibnamefont {Nakano}}, \bibinfo {author}
  {\bibfnamefont {K.}~\bibnamefont {Yakushi}}, \bibinfo {author} {\bibfnamefont
  {S.}~\bibnamefont {Yamada}}, \bibinfo {author} {\bibfnamefont
  {T.}~\bibnamefont {Hiejima}}, \ and\ \bibinfo {author} {\bibfnamefont
  {A.}~\bibnamefont {Kawamoto}},\ }\href
  {https://link.aps.org/doi/10.1103/PhysRevB.82.075134} {\bibfield  {journal}
  {\bibinfo  {journal} {Phys. Rev. B}\ }\textbf {\bibinfo {volume} {82}},\
  \bibinfo {pages} {75134} (\bibinfo {year} {2010})}\BibitemShut {NoStop}%
\bibitem [{\citenamefont {Ivek}\ \emph {et~al.}(2011)\citenamefont {Ivek},
  \citenamefont {Korin-Hamzi{\'{c}}}, \citenamefont {Milat}, \citenamefont
  {Tomi{\'{c}}}, \citenamefont {Clauss}, \citenamefont {Drichko}, \citenamefont
  {Schweitzer},\ and\ \citenamefont {Dressel}}]{Ivek2011}%
  \BibitemOpen
  \bibfield  {author} {\bibinfo {author} {\bibfnamefont {T.}~\bibnamefont
  {Ivek}}, \bibinfo {author} {\bibfnamefont {B.}~\bibnamefont
  {Korin-Hamzi{\'{c}}}}, \bibinfo {author} {\bibfnamefont {O.}~\bibnamefont
  {Milat}}, \bibinfo {author} {\bibfnamefont {S.}~\bibnamefont {Tomi{\'{c}}}},
  \bibinfo {author} {\bibfnamefont {C.}~\bibnamefont {Clauss}}, \bibinfo
  {author} {\bibfnamefont {N.}~\bibnamefont {Drichko}}, \bibinfo {author}
  {\bibfnamefont {D.}~\bibnamefont {Schweitzer}}, \ and\ \bibinfo {author}
  {\bibfnamefont {M.}~\bibnamefont {Dressel}},\ }\href {\doibase
  10.1103/PhysRevB.83.165128} {\bibfield  {journal} {\bibinfo  {journal} {Phys.
  Rev. B}\ }\textbf {\bibinfo {volume} {83}},\ \bibinfo {pages} {165128}
  (\bibinfo {year} {2011})}\BibitemShut {NoStop}%
\bibitem [{Note3()}]{Note3}%
  \BibitemOpen
  \bibinfo {note} {The spectrum is actually comprised of a superposition of 8
  inequivalent but unresolved protons sites. See Supplemental Material for
  angle-dependent measurements.}\BibitemShut {Stop}%
\bibitem [{\citenamefont {Johnston}(2006)}]{Johnston2006}%
  \BibitemOpen
  \bibfield  {author} {\bibinfo {author} {\bibfnamefont {D.~C.}\ \bibnamefont
  {Johnston}},\ }\href {\doibase 10.1103/PhysRevB.74.184430} {\bibfield
  {journal} {\bibinfo  {journal} {Phys. Rev. B}\ }\textbf {\bibinfo {volume}
  {74}},\ \bibinfo {pages} {184430} (\bibinfo {year} {2006})}\BibitemShut
  {NoStop}%
\bibitem [{\citenamefont {Itou}\ \emph {et~al.}(2010)\citenamefont {Itou},
  \citenamefont {Oyamada}, \citenamefont {Maegawa},\ and\ \citenamefont
  {Kato}}]{Itou2010}%
  \BibitemOpen
  \bibfield  {author} {\bibinfo {author} {\bibfnamefont {T.}~\bibnamefont
  {Itou}}, \bibinfo {author} {\bibfnamefont {A.}~\bibnamefont {Oyamada}},
  \bibinfo {author} {\bibfnamefont {S.}~\bibnamefont {Maegawa}}, \ and\
  \bibinfo {author} {\bibfnamefont {R.}~\bibnamefont {Kato}},\ }\href
  {http://dx.doi.org/10.1038/nphys1715} {\bibfield  {journal} {\bibinfo
  {journal} {Nat. Phys.}\ }\textbf {\bibinfo {volume} {6}},\ \bibinfo {pages}
  {673} (\bibinfo {year} {2010})}\BibitemShut {NoStop}%
\bibitem [{\citenamefont {Poirier}\ \emph {et~al.}(2012)\citenamefont
  {Poirier}, \citenamefont {Parent}, \citenamefont {C{\^{o}}t{\'{e}}},
  \citenamefont {Miyagawa}, \citenamefont {Kanoda},\ and\ \citenamefont
  {Shimizu}}]{Poirier2012}%
  \BibitemOpen
  \bibfield  {author} {\bibinfo {author} {\bibfnamefont {M.}~\bibnamefont
  {Poirier}}, \bibinfo {author} {\bibfnamefont {S.}~\bibnamefont {Parent}},
  \bibinfo {author} {\bibfnamefont {A.}~\bibnamefont {C{\^{o}}t{\'{e}}}},
  \bibinfo {author} {\bibfnamefont {K.}~\bibnamefont {Miyagawa}}, \bibinfo
  {author} {\bibfnamefont {K.}~\bibnamefont {Kanoda}}, \ and\ \bibinfo {author}
  {\bibfnamefont {Y.}~\bibnamefont {Shimizu}},\ }\href {\doibase
  10.1103/PhysRevB.85.134444} {\bibfield  {journal} {\bibinfo  {journal} {Phys.
  Rev. B}\ }\textbf {\bibinfo {volume} {85}},\ \bibinfo {pages} {134444}
  (\bibinfo {year} {2012})}\BibitemShut {NoStop}%
\bibitem [{Note4()}]{Note4}%
  \BibitemOpen
  \bibinfo {note} {Upon proper renormalization to the gyromagnetic ratios and
  the local charge density, cf. Fig.~4(a) in Ref.~\protect \rev@citealp
  {Shimizu2016}, the $^1$H and $^{13}$C data coincide in the relevant
  temperature range ($k_B T\leq J$) when acquired at the same magnetic field.
  Upon changing $B_0$, the apparently intrinsic response at temperatures above
  the maximum remains unaltered, while the low-$T$ behavior exhibits pronounced
  modifications, including suppression by field \cite
  {Shimizu2016}.}\BibitemShut {Stop}%
\bibitem [{\citenamefont {Pratt}\ \emph {et~al.}(2011)\citenamefont {Pratt},
  \citenamefont {Baker}, \citenamefont {Blundell}, \citenamefont {Lancaster},
  \citenamefont {Ohira-Kawamura}, \citenamefont {Baines}, \citenamefont
  {Shimizu}, \citenamefont {Kanoda}, \citenamefont {Watanabe},\ and\
  \citenamefont {Saito}}]{Pratt2011}%
  \BibitemOpen
  \bibfield  {author} {\bibinfo {author} {\bibfnamefont {F.~L.}\ \bibnamefont
  {Pratt}}, \bibinfo {author} {\bibfnamefont {P.~J.}\ \bibnamefont {Baker}},
  \bibinfo {author} {\bibfnamefont {S.~J.}\ \bibnamefont {Blundell}}, \bibinfo
  {author} {\bibfnamefont {T.}~\bibnamefont {Lancaster}}, \bibinfo {author}
  {\bibfnamefont {S.}~\bibnamefont {Ohira-Kawamura}}, \bibinfo {author}
  {\bibfnamefont {C.}~\bibnamefont {Baines}}, \bibinfo {author} {\bibfnamefont
  {Y.}~\bibnamefont {Shimizu}}, \bibinfo {author} {\bibfnamefont
  {K.}~\bibnamefont {Kanoda}}, \bibinfo {author} {\bibfnamefont
  {I.}~\bibnamefont {Watanabe}}, \ and\ \bibinfo {author} {\bibfnamefont
  {G.}~\bibnamefont {Saito}},\ }\href {https://doi.org/10.1038/nature09910
  http://10.0.4.14/nature09910
  https://www.nature.com/articles/nature09910{\#}supplementary-information}
  {\bibfield  {journal} {\bibinfo  {journal} {Nature}\ }\textbf {\bibinfo
  {volume} {471}},\ \bibinfo {pages} {612} (\bibinfo {year}
  {2011})}\BibitemShut {NoStop}%
\bibitem [{\citenamefont {Isono}\ \emph {et~al.}(2016)\citenamefont {Isono},
  \citenamefont {Terashima}, \citenamefont {Miyagawa}, \citenamefont {Kanoda},\
  and\ \citenamefont {Uji}}]{Isono2016}%
  \BibitemOpen
  \bibfield  {author} {\bibinfo {author} {\bibfnamefont {T.}~\bibnamefont
  {Isono}}, \bibinfo {author} {\bibfnamefont {T.}~\bibnamefont {Terashima}},
  \bibinfo {author} {\bibfnamefont {K.}~\bibnamefont {Miyagawa}}, \bibinfo
  {author} {\bibfnamefont {K.}~\bibnamefont {Kanoda}}, \ and\ \bibinfo {author}
  {\bibfnamefont {S.}~\bibnamefont {Uji}},\ }\href {\doibase
  10.1038/ncomms13494} {\bibfield  {journal} {\bibinfo  {journal} {Nat.
  Commun.}\ }\textbf {\bibinfo {volume} {7}},\ \bibinfo {pages} {13494}
  (\bibinfo {year} {2016})}\BibitemShut {NoStop}%
\bibitem [{\citenamefont {Isono}\ \emph {et~al.}(2018)\citenamefont {Isono},
  \citenamefont {Sugiura}, \citenamefont {Terashima}, \citenamefont {Miyagawa},
  \citenamefont {Kanoda},\ and\ \citenamefont {Uji}}]{Isono2018}%
  \BibitemOpen
  \bibfield  {author} {\bibinfo {author} {\bibfnamefont {T.}~\bibnamefont
  {Isono}}, \bibinfo {author} {\bibfnamefont {S.}~\bibnamefont {Sugiura}},
  \bibinfo {author} {\bibfnamefont {T.}~\bibnamefont {Terashima}}, \bibinfo
  {author} {\bibfnamefont {K.}~\bibnamefont {Miyagawa}}, \bibinfo {author}
  {\bibfnamefont {K.}~\bibnamefont {Kanoda}}, \ and\ \bibinfo {author}
  {\bibfnamefont {S.}~\bibnamefont {Uji}},\ }\href {\doibase
  10.1038/s41467-018-04005-1} {\bibfield  {journal} {\bibinfo  {journal} {Nat.
  Commun.}\ }\textbf {\bibinfo {volume} {9}},\ \bibinfo {pages} {1509}
  (\bibinfo {year} {2018})}\BibitemShut {NoStop}%
\bibitem [{\citenamefont {{\'{S}}wietlik}\ \emph {et~al.}(2017)\citenamefont
  {{\'{S}}wietlik}, \citenamefont {Barszcz}, \citenamefont {Pustogow},\ and\
  \citenamefont {Dressel}}]{Swietlik2017}%
  \BibitemOpen
  \bibfield  {author} {\bibinfo {author} {\bibfnamefont {R.}~\bibnamefont
  {{\'{S}}wietlik}}, \bibinfo {author} {\bibfnamefont {B.}~\bibnamefont
  {Barszcz}}, \bibinfo {author} {\bibfnamefont {A.}~\bibnamefont {Pustogow}}, \
  and\ \bibinfo {author} {\bibfnamefont {M.}~\bibnamefont {Dressel}},\ }\href
  {https://link.aps.org/doi/10.1103/PhysRevB.95.085205} {\bibfield  {journal}
  {\bibinfo  {journal} {Phys. Rev. B}\ }\textbf {\bibinfo {volume} {95}},\
  \bibinfo {pages} {85205} (\bibinfo {year} {2017})}\BibitemShut {NoStop}%
\bibitem [{\citenamefont {Padmalekha}\ \emph {et~al.}(2015)\citenamefont
  {Padmalekha}, \citenamefont {Blankenhorn}, \citenamefont {Ivek},
  \citenamefont {Bogani}, \citenamefont {Schlueter},\ and\ \citenamefont
  {Dressel}}]{Padmalekha2015}%
  \BibitemOpen
  \bibfield  {author} {\bibinfo {author} {\bibfnamefont {K.~G.}\ \bibnamefont
  {Padmalekha}}, \bibinfo {author} {\bibfnamefont {M.}~\bibnamefont
  {Blankenhorn}}, \bibinfo {author} {\bibfnamefont {T.}~\bibnamefont {Ivek}},
  \bibinfo {author} {\bibfnamefont {L.}~\bibnamefont {Bogani}}, \bibinfo
  {author} {\bibfnamefont {J.~A.}\ \bibnamefont {Schlueter}}, \ and\ \bibinfo
  {author} {\bibfnamefont {M.}~\bibnamefont {Dressel}},\ }\href {\doibase
  https://doi.org/10.1016/j.physb.2014.11.073} {\bibfield  {journal} {\bibinfo
  {journal} {Physica B: Condens. Matter}\ }\textbf {\bibinfo {volume} {460}},\
  \bibinfo {pages} {211} (\bibinfo {year} {2015})}\BibitemShut {NoStop}%
\bibitem [{\citenamefont {Bourgeois-Hope}\ \emph {et~al.}(2019)\citenamefont
  {Bourgeois-Hope}, \citenamefont {Lalibert{\'{e}}}, \citenamefont
  {Lefran{\c{c}}ois}, \citenamefont {Grissonnanche}, \citenamefont {{Ren{\'{e}}
  de Cotret}}, \citenamefont {Gordon}, \citenamefont {Kitou}, \citenamefont
  {Sawa}, \citenamefont {Cui}, \citenamefont {Kato}, \citenamefont
  {Taillefer},\ and\ \citenamefont {Doiron-Leyraud}}]{Bourgeois-Hope2019}%
  \BibitemOpen
  \bibfield  {author} {\bibinfo {author} {\bibfnamefont {P.}~\bibnamefont
  {Bourgeois-Hope}}, \bibinfo {author} {\bibfnamefont {F.}~\bibnamefont
  {Lalibert{\'{e}}}}, \bibinfo {author} {\bibfnamefont {E.}~\bibnamefont
  {Lefran{\c{c}}ois}}, \bibinfo {author} {\bibfnamefont {G.}~\bibnamefont
  {Grissonnanche}}, \bibinfo {author} {\bibfnamefont {S.}~\bibnamefont
  {{Ren{\'{e}} de Cotret}}}, \bibinfo {author} {\bibfnamefont {R.}~\bibnamefont
  {Gordon}}, \bibinfo {author} {\bibfnamefont {S.}~\bibnamefont {Kitou}},
  \bibinfo {author} {\bibfnamefont {H.}~\bibnamefont {Sawa}}, \bibinfo {author}
  {\bibfnamefont {H.}~\bibnamefont {Cui}}, \bibinfo {author} {\bibfnamefont
  {R.}~\bibnamefont {Kato}}, \bibinfo {author} {\bibfnamefont {L.}~\bibnamefont
  {Taillefer}}, \ and\ \bibinfo {author} {\bibfnamefont {N.}~\bibnamefont
  {Doiron-Leyraud}},\ }\href@noop {} {\  (\bibinfo {year} {2019})},\ \Eprint
  {http://arxiv.org/abs/1904.10402} {arXiv:1904.10402} \BibitemShut {NoStop}%
\end{thebibliography}

\begin{thebibliography}{8}%
\makeatletter
\providecommand \@ifxundefined [1]{%
 \@ifx{#1\undefined}
}%
\providecommand \@ifnum [1]{%
 \ifnum #1\expandafter \@firstoftwo
 \else \expandafter \@secondoftwo
 \fi
}%
\providecommand \@ifx [1]{%
 \ifx #1\expandafter \@firstoftwo
 \else \expandafter \@secondoftwo
 \fi
}%
\providecommand \natexlab [1]{#1}%
\providecommand \enquote  [1]{``#1''}%
\providecommand \bibnamefont  [1]{#1}%
\providecommand \bibfnamefont [1]{#1}%
\providecommand \citenamefont [1]{#1}%
\providecommand \href@noop [0]{\@secondoftwo}%
\providecommand \href [0]{\begingroup \@sanitize@url \@href}%
\providecommand \@href[1]{\@@startlink{#1}\@@href}%
\providecommand \@@href[1]{\endgroup#1\@@endlink}%
\providecommand \@sanitize@url [0]{\catcode `\\12\catcode `\$12\catcode
  `\&12\catcode `\#12\catcode `\^12\catcode `\_12\catcode `\%12\relax}%
\providecommand \@@startlink[1]{}%
\providecommand \@@endlink[0]{}%
\providecommand \url  [0]{\begingroup\@sanitize@url \@url }%
\providecommand \@url [1]{\endgroup\@href {#1}{\urlprefix }}%
\providecommand \urlprefix  [0]{URL }%
\providecommand \Eprint [0]{\href }%
\providecommand \doibase [0]{http://dx.doi.org/}%
\providecommand \selectlanguage [0]{\@gobble}%
\providecommand \bibinfo  [0]{\@secondoftwo}%
\providecommand \bibfield  [0]{\@secondoftwo}%
\providecommand \translation [1]{[#1]}%
\providecommand \BibitemOpen [0]{}%
\providecommand \bibitemStop [0]{}%
\providecommand \bibitemNoStop [0]{.\EOS\space}%
\providecommand \EOS [0]{\spacefactor3000\relax}%
\providecommand \BibitemShut  [1]{\csname bibitem#1\endcsname}%
\let\auto@bib@innerbib\@empty
\bibitem [S1]{SGati2018a}%
  \BibitemOpen
  \bibfield  {author} {\bibinfo {author} {\bibfnamefont {E.}~\bibnamefont
  {Gati}}, \bibinfo {author} {\bibfnamefont {J.~K.~H.}\ \bibnamefont
  {Fischer}}, \bibinfo {author} {\bibfnamefont {P.}~\bibnamefont
  {Lunkenheimer}}, \bibinfo {author} {\bibfnamefont {D.}~\bibnamefont
  {Zielke}}, \bibinfo {author} {\bibfnamefont {S.}~\bibnamefont
  {K{\"{o}}hler}}, \bibinfo {author} {\bibfnamefont {F.}~\bibnamefont {Kolb}},
  \bibinfo {author} {\bibfnamefont {H.-A.~K.}\ \bibnamefont {von Nidda}},
  \bibinfo {author} {\bibfnamefont {S.~M.}\ \bibnamefont {Winter}}, \bibinfo
  {author} {\bibfnamefont {H.}~\bibnamefont {Schubert}}, \bibinfo {author}
  {\bibfnamefont {J.~A.}\ \bibnamefont {Schlueter}}, \bibinfo {author}
  {\bibfnamefont {H.~O.}\ \bibnamefont {Jeschke}}, \bibinfo {author}
  {\bibfnamefont {R.}~\bibnamefont {Valent{\'{i}}}}, \ and\ \bibinfo {author}
  {\bibfnamefont {M.}~\bibnamefont {Lang}},\ }\href {\doibase
  10.1103/PhysRevLett.120.247601} {\bibfield  {journal} {\bibinfo  {journal}
  {Phys. Rev. Lett.}\ }\textbf {\bibinfo {volume} {120}},\ \bibinfo {pages}
  {247601} (\bibinfo {year} {2018}{\natexlab{a}})}\BibitemShut {NoStop}%
\bibitem [S2]{SShimizu2003}%
  \BibitemOpen
  \bibfield  {author} {\bibinfo {author} {\bibfnamefont {Y.}~\bibnamefont
  {Shimizu}}, \bibinfo {author} {\bibfnamefont {K.}~\bibnamefont {Miyagawa}},
  \bibinfo {author} {\bibfnamefont {K.}~\bibnamefont {Kanoda}}, \bibinfo
  {author} {\bibfnamefont {M.}~\bibnamefont {Maesato}}, \ and\ \bibinfo
  {author} {\bibfnamefont {G.}~\bibnamefont {Saito}},\ }\href
  {https://link.aps.org/doi/10.1103/PhysRevLett.91.107001} {\bibfield
  {journal} {\bibinfo  {journal} {Phys. Rev. Lett.}\ }\textbf {\bibinfo
  {volume} {91}},\ \bibinfo {pages} {107001} (\bibinfo {year}
  {2003})}\BibitemShut {NoStop}%
\bibitem [S3]{SDrichko2014}%
  \BibitemOpen
  \bibfield  {author} {\bibinfo {author} {\bibfnamefont {N.}~\bibnamefont
  {Drichko}}, \bibinfo {author} {\bibfnamefont {R.}~\bibnamefont {Beyer}},
  \bibinfo {author} {\bibfnamefont {E.}~\bibnamefont {Rose}}, \bibinfo {author}
  {\bibfnamefont {M.}~\bibnamefont {Dressel}}, \bibinfo {author} {\bibfnamefont
  {J.~A.}\ \bibnamefont {Schlueter}}, \bibinfo {author} {\bibfnamefont {S.~A.}\
  \bibnamefont {Turunova}}, \bibinfo {author} {\bibfnamefont {E.~I.}\
  \bibnamefont {Zhilyaeva}}, \ and\ \bibinfo {author} {\bibfnamefont {R.~N.}\
  \bibnamefont {Lyubovskaya}},\ }\href
  {https://link.aps.org/doi/10.1103/PhysRevB.89.075133} {\bibfield  {journal}
  {\bibinfo  {journal} {Phys. Rev. B}\ }\textbf {\bibinfo {volume} {89}},\
  \bibinfo {pages} {75133} (\bibinfo {year} {2014})}\BibitemShut {NoStop}%
\bibitem [S4]{SShimizu2016}%
  \BibitemOpen
  \bibfield  {author} {\bibinfo {author} {\bibfnamefont {Y.}~\bibnamefont
  {Shimizu}}, \bibinfo {author} {\bibfnamefont {T.}~\bibnamefont {Hiramatsu}},
  \bibinfo {author} {\bibfnamefont {M.}~\bibnamefont {Maesato}}, \bibinfo
  {author} {\bibfnamefont {A.}~\bibnamefont {Otsuka}}, \bibinfo {author}
  {\bibfnamefont {H.}~\bibnamefont {Yamochi}}, \bibinfo {author} {\bibfnamefont
  {A.}~\bibnamefont {Ono}}, \bibinfo {author} {\bibfnamefont {M.}~\bibnamefont
  {Itoh}}, \bibinfo {author} {\bibfnamefont {M.}~\bibnamefont {Yoshida}},
  \bibinfo {author} {\bibfnamefont {M.}~\bibnamefont {Takigawa}}, \bibinfo
  {author} {\bibfnamefont {Y.}~\bibnamefont {Yoshida}}, \ and\ \bibinfo
  {author} {\bibfnamefont {G.}~\bibnamefont {Saito}},\ }\href
  {https://link.aps.org/doi/10.1103/PhysRevLett.117.107203} {\bibfield
  {journal} {\bibinfo  {journal} {Phys. Rev. Lett.}\ }\textbf {\bibinfo
  {volume} {117}},\ \bibinfo {pages} {107203} (\bibinfo {year}
  {2016})}\BibitemShut {NoStop}%
\bibitem [S5]{SJeschke2012}%
  \BibitemOpen
  \bibfield  {author} {\bibinfo {author} {\bibfnamefont {H.~O.}\ \bibnamefont
  {Jeschke}}, \bibinfo {author} {\bibfnamefont {M.}~\bibnamefont {de~Souza}},
  \bibinfo {author} {\bibfnamefont {R.}~\bibnamefont {Valent{\'{i}}}}, \bibinfo
  {author} {\bibfnamefont {R.~S.}\ \bibnamefont {Manna}}, \bibinfo {author}
  {\bibfnamefont {M.}~\bibnamefont {Lang}}, \ and\ \bibinfo {author}
  {\bibfnamefont {J.~A.}\ \bibnamefont {Schlueter}},\ }\href
  {https://link.aps.org/doi/10.1103/PhysRevB.85.035125} {\bibfield  {journal}
  {\bibinfo  {journal} {Phys. Rev. B}\ }\textbf {\bibinfo {volume} {85}},\
  \bibinfo {pages} {35125} (\bibinfo {year} {2012})}\BibitemShut {NoStop}%
\bibitem [S6]{SHiramatsu2017}%
  \BibitemOpen
  \bibfield  {author} {\bibinfo {author} {\bibfnamefont {T.}~\bibnamefont
  {Hiramatsu}}, \bibinfo {author} {\bibfnamefont {Y.}~\bibnamefont {Yoshida}},
  \bibinfo {author} {\bibfnamefont {G.}~\bibnamefont {Saito}}, \bibinfo
  {author} {\bibfnamefont {A.}~\bibnamefont {Otsuka}}, \bibinfo {author}
  {\bibfnamefont {H.}~\bibnamefont {Yamochi}}, \bibinfo {author} {\bibfnamefont
  {M.}~\bibnamefont {Maesato}}, \bibinfo {author} {\bibfnamefont
  {Y.}~\bibnamefont {Shimizu}}, \bibinfo {author} {\bibfnamefont
  {H.}~\bibnamefont {Ito}}, \bibinfo {author} {\bibfnamefont {Y.}~\bibnamefont
  {Nakamura}}, \bibinfo {author} {\bibfnamefont {H.}~\bibnamefont {Kishida}},
  \bibinfo {author} {\bibfnamefont {M.}~\bibnamefont {Watanabe}}, \ and\
  \bibinfo {author} {\bibfnamefont {R.}~\bibnamefont {Kumai}},\ }\href
  {\doibase 10.1246/bcsj.20170167} {\bibfield  {journal} {\bibinfo  {journal}
  {Bull. Chem. Soc. Jpn.}\ }\textbf {\bibinfo {volume} {90}},\ \bibinfo {pages}
  {1073} (\bibinfo {year} {2017})}\BibitemShut {NoStop}%
\bibitem [S7]{SAbragam1983}%
  \BibitemOpen
  \bibfield  {author} {\bibinfo {author} {\bibfnamefont {A.}~\bibnamefont
  {Abragam}},\ }\href@noop {} {\emph {\bibinfo {title} {{Principles of Nuclear
  Magnetism}}}}\ (\bibinfo  {publisher} {Oxford University Press},\ \bibinfo
  {address} {Hong Kong},\ \bibinfo {year} {1983})\BibitemShut {NoStop}%
\bibitem [S8]{SGati2018}%
  \BibitemOpen
  \bibfield  {author} {\bibinfo {author} {\bibfnamefont {E.}~\bibnamefont
  {Gati}}, \bibinfo {author} {\bibfnamefont {S.~M.}\ \bibnamefont {Winter}},
  \bibinfo {author} {\bibfnamefont {J.~A.}\ \bibnamefont {Schlueter}}, \bibinfo
  {author} {\bibfnamefont {H.}~\bibnamefont {Schubert}}, \bibinfo {author}
  {\bibfnamefont {J.}~\bibnamefont {M{\"{u}}ller}}, \ and\ \bibinfo {author}
  {\bibfnamefont {M.}~\bibnamefont {Lang}},\ }\href {\doibase
  10.1103/PhysRevB.97.075115} {\bibfield  {journal} {\bibinfo  {journal} {Phys.
  Rev. B}\ }\textbf {\bibinfo {volume} {97}},\ \bibinfo {pages} {75115}
  (\bibinfo {year} {2018}{\natexlab{b}})}\BibitemShut {NoStop}%
\end{thebibliography}
\end{document}